\documentclass[epj]{svjour}
\usepackage{graphics}
\usepackage{amssymb}
\usepackage{bm}
\begin{document}
\title{On the two remaning issues in the gauge-invariant decomposition
problem of the nucleon spin}
\author{M. Wakamatsu\inst{1}
}                     
\offprints{}          
%
\institute{Department of Physics, Faculty of Science,
Osaka University, Toyonaka, Osaka 560-0043, JAPAN}
\date{Received: date / Revised version: date}
%
\abstract{The question whether the total angular momentum of the gluon
in the nucleon can be decomposed into its spin and orbital parts without
conflict with the gauge-invariance principle has been an object of long-lasting
debate. Despite a remarkable progress achieved through the recent intensive
researches, the following two issues still remains to be clarified more
transparently. The first issue is to resolve the apparent conflict between
the proposed gauge-invariant decomposition of the total gluon angular
momentum and the textbook statement that the total angular momentum
of the photon cannot be gauge-invariantly decomposed into its spin
and orbital parts. We show that this problem is also intimately connected
with the uniqueness or non-uniqueness problem of the nucleon spin
decomposition. The second practically more important issue is that,
among the two physically inequivalent
decompositions of the nucleon spin, i.e. the ``canonical''
type decomposition and the ``mechanical'' type decomposition, which
can we say is more physical or closer to direct observation ?
In the present paper, we try to answer both these questions as clearly
as possible. 
} 
\maketitle
\section{Introduction}
\label{intro}

To get a complete decomposition of nucleon spin is a fundamentally important
homework of QCD \cite{EMC88,EMC89}. 
Unfortunately, this is an extremely delicate and difficult problem,
which has been rejecting a clear answer for more than 20 years since
the first pioneering theoretical
consideration in the paper by Jaffe and Manohar \cite{JM90}.
(To overview controversial status of the problem, see two recent
reviews \cite{Review_LL14,Review_Waka14}.)
The central issue here is whether the total gluon
angular momentum can be gauge-invariantly decomposed into its spin and
orbital parts. It has long been believed that the answer is
no \cite{Ji97,Ji98}, just because
many textbooks of electrodynamics clearly state that the total photon
angular momentum cannot be gauge-invariantly decomposed into its spin
and orbital parts \cite{Book_AB65,Book_JR76,Book_BLP82,Book_CDG89}.
On the other hand, we also know the fact that several
experiments with use of paraxial laser beam confirmed that the spin
and the orbital angular momentum of a photon can separately be
measured \cite{Beth36,ABSW92}.
The theoretical basis of  this separation is the familiar
transverse-longitudinal decomposition of the photon field. 
Some years ago, motivated by the idea of transverse-longitudinal
decomposition of the photon field, Chen et al. proposed the idea to
decompose the gluon field into the physical and pure-gauge
component \cite{Chen08,Chen09}.
This enables them to get a gauge-invariant complete decomposition of the
nucleon spin into four pieces, i.e. the spin and orbital angular
momentum (OAM) parts of quarks and the spin and OAM parts of gluons.
Although this is certainly a gauge-invariant decomposition of the nucleon
spin, it is important to recognize that their decomposition
of the gluon field is essentially the standard transverse-longitudinal
decomposition, which is given only after fixing the Lorentz frame of
reference.
Another point is that, even within the framework of this
transverse-longitudinal decomposition,
the way of gauge-invariant decomposition of the nucleon spin is not unique.
In fact, it was pointed out in \cite{Waka10} that another gauge-invariant
complete decomposition of the nucleon spin exists.
The difference with the original decomposition by Chen et al. is
characterized by the OAM parts of quarks and gluons.
The quark and gluon OAMs in the Chen decomposition is essentially the
``canonical'' OAMs, although gauge-invariantized. On the other hand,
the quark and gluon OAMs in the decomposition proposed in \cite{Waka10}
is the gauge-invariant ``mechanical'' OAMs.
It is then reasonable to call these two decompositions the ``canonical'' and
``mechanical'' decompositions of the nucleon spin, 
respectively \cite{Review_LL14,Review_Waka14}.

In either case, since these two decompositions are given in a fixed Lorentz
frame, or given in noncovariant forms, it is not very convenient when
investigating the relations with high-energy deep-inelastic-scattering (DIS)
observables, which is the field of physics where the problem of the
nucleon spin decomposition came into existence.
In the paper \cite{Waka11A}, we therefore proposed more general form of
nucleon spin decomposition, which has ``seemingly'' covariant appearance.
It was shown that, based only upon a few general conditions, one can get
two different ``seemingly'' covariant forms of gauge-invariant
complete decompositions of the nucleon spin,
which was called in \cite{Waka11A} the decomposition (I) and (II).
Formally, these two provides us with the most general gauge-invariant
decomposition of the nucleon spin, in the sense that any gauge-invariant
decomposition falls into either of the two. In fact,
the decomposition (II) was verified to contain the gauge-invariant
Bashinsky-Jaffe decomposition \cite{BJ99} motivated by the light-cone
gauge.
It was also shown to contain the original Chen decomposition
after an appropriate choice of the Lorentz frame. They both fall into the
category of ``canonical'' decomposition. On the other hand, another
decomposition (I) falls into the category of ``mechanical'' decomposition.
It contains the noncovariant decomposition given in \cite{Waka10}
after fixing the Lorentz frame, but it is more general \cite{Waka12}.

Although most inclusive, the problem of this general framework is that
the decomposition of the gluon field into the physical and pure-gauge
components has large freedom or arbitrariness.
In fact, it was criticized by several
researchers that our formal decomposition of the gauge field into its
physical and pure-gauge components is not unique at all and there
can be infinitely many such decompositions, thereby being led them to the
conclusion that there are in principle infinitely many decomposition of
the nucleon spin \cite{JXY12,JXZ12,Lorce13A,Lorce13B,Lorce14}.    
This uniqueness or non-uniqueness problem of the nucleon spin decomposition
is one of the two remaining issues of the gauge-invariant
decomposition problem of the nucleon spin summarized as

\begin{enumerate}
\item Are there infinitely many decompositions of the nucleon
spin \cite{Waka13} ?
If not, what physical principle favors one particular
decomposition among many candidates ? 

\item Among the two decompositions, i.e. the ``canonical'' type
decomposition and the ``mechanical'' type decomposition, which can we
say is more physical ? More ``physical'' here means that it is
closer to direct measurements.
\end{enumerate} 

The purpose of the present paper is to answer these questions as convincingly
as possible. First, in sect.2, we show that the first problem is inseparably
connected with our original question, i.e. the apparent conflict between
the proposed gauge-invariant decomposition of the nucleon spin, especially
the gluon part, and the textbook statement that the total angular momentum
of the photon cannot be gauge-invariantly decomposed into the spin and
orbital parts. Actually, the conflict can be avoided in the following way.
First, more precise statement on the decomposition of the total photon
angular momentum should be the following. The total angular momentum
of the photon cannot be decomposed into its spin and orbital parts
so as to meet both the requirements of gauge-invariance and of the
Lorentz-frame independence, simultaneously. Once the Lorentz frame
is fixed, the decomposition can be made gauge-invariantly without
any difficulty. This already indicates that what makes the decomposition
problem of the total gluon angular momentum in the nucleon difficult is
an intricate interplay between the gauge and Lorentz symmetry.
In fact, we shall show that the key factor, which uniquely select one particular
decomposition from many possible gauge-invariant decompositions of the
nucleon spin, is the Lorentz-boost-invariance in the momentum direction
of the parent nucleon.

Next, in sect.3, we discuss the second issue. 
It is known that the average orbital angular momentum (OAM) of quarks
along the nucleon momentum defined by the Wigner distribution
coincides with the gauge-invariant canonical OAM not the mechanical
(or kinetic) OAM. There is also a claim that the average transverse
momentum of quarks defined through the Wigner distribution coincides
with the ``canonical'' momentum not the ``mechanical'' momentum.
We shall show that the latter claim is not necessarily true and then give a
universally correct physical interpretation of the average momentum and
OAM defined by the Wigner distribution. With the help of these considerations,
we also carry out a detailed comparative analysis of the canonical and mechanical
OAMs, by putting special emphasis upon the relation with direct
deep-inelastic-scattering (DIS) observables. Summary and conclusion are
then given in sect.4.

\section{The role of Lorentz symmetry in the gauge-invariant decomposition
problem of the nucleon spin}
\label{sec:1}

It is very important to clearly recognize that the transverse-longitudinal
decomposition is given only noncovariantly, i.e. only after fixing a Lorentz-frame
of reference.
Seemingly covariant extension of the gauge-invariant decomposition is then
proposed in the paper \cite{Waka11A}.
As expected from already known decompositions given
in noncovariant forms \cite{Waka10}, we naturally get two ``seemingly''
covariant decompositions of the QCD angular momentum tensor, which are
physically inequivalent.
(The word ``seemingly'' is important here, because the decomposition of the
gauge field into the physical and pure-gauge
component is intrinsically noncovariant even though the appearance looks
covariant.)
The one is the ``canonical'' type decomposition given as
\begin{eqnarray}
 M^{\lambda \mu \nu}_{QCD} &=&  
 M^{\prime \lambda \mu \nu}_{q-spin} +  
 M^{\prime \lambda \mu \nu}_{q-OAM} +   
 M^{\prime \lambda \mu \nu}_{G-spin} +   
 M^{\prime \lambda \mu \nu}_{G-OAM} \nonumber \\ 
 &+& \ \mbox{\rm boost} \ + \ \mbox{\rm total divergence} ,
 \label{Eq:can_decomp}
\end{eqnarray}
where
\begin{eqnarray}
 M^{\prime \lambda \mu \nu}_{q-spin}
 &=& \frac{1}{2} \,\epsilon^{\lambda \mu \nu \sigma} \,
 \bar{\psi} \,\gamma_{\sigma} \,\gamma_5 \,\psi , \\
 M^{\prime \lambda \mu \nu}_{q-OAM} 
 &=& \bar{\psi} \,\gamma^{\lambda} \,(\,x^{\mu} \,i \,
 D^{\nu}_{pure}  
 -  x^{\nu} \,i \,
 D^{\mu}_{pure} \,) \,\psi , \\
 M^{\prime \lambda \mu \nu}_{G-spin}
 &=& 2 \,\mbox{Tr} \,\{\, F^{\lambda \nu} \,A_{phys}^{\mu} \,
 -  F^{\lambda \mu} \,A_{phys}^{\nu} \,\} , \\
 M^{\prime \lambda \mu \nu}_{G-OAM}
 &=& 2 \,\mbox{Tr} \,\{ F^{\lambda \alpha} \,
 (x^{\mu} \,{\cal D}_{pure}^{\nu}
 - x^{\nu} \,{\cal D}_{pure}^{\mu} ) \,A_{\alpha}^{phys} \} .\ \ \ 
\end{eqnarray}
Here, $D^\mu_{pure} = \partial^\mu - \,i \,g \,A^\mu_{pure}$ is the
pure-gauge covariant derivative for the fundamental representation of color
SU(3), while ${\cal D}^\mu_{pure} = \partial - i \,g \,[A^\mu_{pure}, \cdot \,]$
is the pure-gauge covariant derivative for the adjoint representation.
The other is the ``mechanical'' type decomposition given as
\begin{eqnarray}
 M^{\lambda \mu \nu}_{QCD} &=&  
 M^{\lambda \mu \nu}_{q - spin}
 + M^{\lambda \mu \nu}_{q - OAM} + 
 M^{\mu \nu \lambda}_{G - spin}
 + M^{\lambda \mu \nu}_{G - OAM} \nonumber \\
 &+& \ \mbox{\rm boost}  
 \ + \ \mbox{total divergence} ,
 \label{Eq:mech_decomp}
\end{eqnarray}
where
\begin{eqnarray}
 M^{\lambda \mu \nu}_{q - spin}
 &=& M^{\prime \lambda \mu \nu}_{q-spin} , \\
 M^{\lambda \mu \nu}_{q - OAM}
 &=& \bar{\psi} \,\gamma^\lambda \,(\,x^{\mu} \,i \,
 D^{\nu}  
 \ - \ x^{\nu} \,i \,
 D^{\mu} \,) \,\psi , \\
 M^{\lambda \mu \nu}_{G - spin} 
 &=& M^{\prime \lambda \mu \nu}_{G-spin} , \\
 M^{\lambda \mu \nu}_{g - OAM} 
 &=& M^{\prime \lambda \mu \nu}_{G-OAM} \nonumber \\
 + \ 2 &\,& \!\!\!\!\!\!\!
 \mbox{Tr} \,[\, (\,{\cal D}_{\alpha} \,F^{\alpha \lambda} \,)
 \,(\,x^{\mu} \, A^{\nu}_{phys} \ - \ 
 x^{\nu} \, A^{\mu}_{phys} \,) \,] , \ \ \ 
\end{eqnarray}
where $D^\mu$ and ${\cal D}^\mu$ are the standard covariant derivative
respectively acting on the fundamental and adjoint representations of color
SU(3). To obtain these ``seemingly'' covariant decompositions, we need to
impose only a few general conditions on the physical and pure-gauge
components of the gluon field. They are the pure-gauge condition for
the pure-gauge component of the gluon field
\begin{eqnarray}
 F^{\mu \nu}_{pure}
 \equiv 
 \partial^\mu \,A^\nu_{pure} - 
 \partial^\nu \,A^\mu_{pure} - i \,g \,
 [A^\mu_{pure}, A^\nu_{pure}] 
 = 0 , \ \ \ \ \ \label{Eq:Gcond1}
\end{eqnarray}
and the homogeneous (covariant) and inhomogeneous gauge transformation
properties for the physical and pure-gauge components : 
\begin{eqnarray}
 A^\mu_{phys}(x) &\rightarrow&
 U(x) \,A^\mu_{phys}(x) \,U^{-1}(x) \label{Eq:Gcond2} , \\
 A^\mu_{pure}(x) &\rightarrow&
 U(x) \,\left(\,A^\mu_{pure}(x) + \frac{i}{g}
 \,\, \partial^\mu \,\right) \,U^{-1}(x) . \ \ \ \label{Eq:Gcond3}
\end{eqnarray}
Actually, these conditions are too generic and they are not enough to fix
the decomposition uniquely.
Nevertheless, it was shown there that one of the above decompositions,
i.e. the ``canonical'' type decomposition (\ref{Eq:can_decomp}), contains the
LC-gauge motivated Bashinsky-Jaffe decomposition \cite{BJ99}
as well as the Chen decomposition \cite{Chen08,Chen09}
as special cases. 
The nonuniqueness of our formal decomposition is criticized by several
researches \cite{JXY12,JXZ12,Lorce13A,Lorce13B,Lorce14}.
This leads them to the conclusion
that there are in principle infinitely many decomposition of the nucleon
spin. According to Ji et al. \cite{JXY12,JXZ12}, the arbitrariness of
the decomposition comes from
the path-dependence of the Wilson line, which is necessary for the decomposition
of the gauge field into the physical and pure-gauge components.
Another argument in favor of the existence of infinitely many decomposition
of the nucleon spin was advocated by Lorce \cite{Lorce13A,Lorce13B,Lorce14},
based on what-he-call the Stueckelberg symmetry, which
changes both of $A_{phys}$ and $A_{pure}$, while leaving their sum intact.
(The idea was first presented in \cite{Stoilov14}. This geometric idea is
being pursued further in a recent paper \cite{FLM14}.) 

In a recent paper \cite{JZZ13}, Ji, Xu, and Zhao stepped further and
showed that the total
gluon helicity in a polarized proton is shown to be large momentum limit of a gauge-invariant operator $\bm{E} \times \bm{A}_\perp$ with $\bm{A}_\perp$ being
the transverse component of the gauge potential.
Their argument goes as follows.                 
First, they pointed out that, for the abelian gluon, 
the gluon spin operator $S_G$, which is deduced from the definition of
the longitudinally polarized gluon distribution as its first moment, 
can be expressed in the following form : 
\begin{eqnarray}
 S_G \ = \ (\bm{E}(0) \times 
 \bm{A}_{phys}(0))^3 ,
\end{eqnarray}
with
\begin{eqnarray}
 \bm{A}_{phys} (0) 
 \ = \ \bm{A} (0) \ - \ 
 \left. \frac{1}{\nabla^+} \,\nabla \,A^+ (\xi^-)
 \right|_{\xi^- = 0} .
\end{eqnarray}
Next, they showed that the above operator is just the IMF
limit of the operator $\bm{E} \times \bm{A}_\perp$.
From this fact, they concluded that, to identify $\bm{E} \times \bm{A}_\perp$
 as the gluon helicity, one must have the following conditions,
that is, the IMF and physical gauge, i.e. the light-cone gauge.
The statement would be nothing wrong, but
it has a danger of causing a little misunderstanding.
In fact, the gluon spin, or the longitudinally polarized gluon distribution,
which appears in the standard formulation of the parton distribution
functions (PDFs), has an important property that it does not depend
on the magnitude of the nucleon momentum.
Since this boost-invariance of general collinear parton distributions (PDFs)
along the direction of the nucleon momentum plays a decisive role
in our subsequent argument, we think it useful to convince this
property following Collins' textbook \cite{Book_Collins11}.
Let us start with the well-known definition of the simplest unpolarized PDF
of the nucleon given as
\begin{equation}
 \int \,\frac{d \lambda}{2 \,\pi} \,e^{\,i \,\lambda \,k \cdot n} \,\,
 \langle P \,|\,\bar{\psi} (0) \not\!n \,\psi (\lambda \,n) \,|\, P \rangle,
 \label{Eq:UnpolPDF}
\end{equation}
with $n$ being the standard light-like vector, while $P$ is the
momentum of the nucleon. Since the r.h.s. is a scalar function,
it must be a function of $k \cdot n$ and $P \cdot n$ as
\begin{equation}
 \tilde{q} \,(k \cdot n, \,P \cdot n) .
\end{equation}
The expression (\ref{Eq:UnpolPDF}) is obviously invariant under the scaling
of the 4-vector $n$ by an arbitrary positive factor, which means that
only the combination $x \equiv k \cdot n \,/\, P \cdot n$ is allowed.
This gives the standard definition of the unpolarized PDF given as
\begin{equation}
 q (x) \ = \ \int \,\frac{d \lambda}{2 \,\pi} \,\,
 e^{\,i \,\lambda \,x \,(P \cdot n)} \,\,
 \langle P \,|\,\bar{\psi} (0) \not\!n \,\psi (\lambda \,n) \,|\, P \rangle .
\end{equation}
Now let us consider the Lorentz boost with a velocity $v$ along the direction
of the nucleon momentum $P$, which we can take the 3-direction without loss
of generality.
It is given by
\begin{eqnarray}
 &\,& x^0 \ \rightarrow \ \gamma \,(x^0 \ - \ v \,x^3), \ \ \ \ 
 x^1 \ \rightarrow \ x^1, \nonumber \\
 &\,& x^2 \ \rightarrow \ x^2, \ \ \ \ 
 x^3 \ \rightarrow \ \gamma \,(x^3 \ - \ v \,x^0), \label{Eq:Boost} \hspace{5mm}
\end{eqnarray}
with $\gamma = 1 \,/\,\sqrt{1 - v^2}$. Under this boost, $k \cdot n$ and
$P \cdot n$ transform as
\begin{equation}
 k \cdot n \ \rightarrow \ \gamma \,(1 - v) \,k \cdot n, \ \ 
 P \cdot n \ \rightarrow \ \gamma \,(1 - v) \,P \cdot n,
\end{equation}
so that the ratio $x = (k \cdot n) \,/\, (P \cdot n)$ is obviously
invariant under it.
Naturally, this also applies to the
longitudinally polarized gluon distributions and also the gluon spin,
which is defined as the first moment of the former.

To see the importance of the constraint of Lorentz-frame independence
in our problem of the gauge-invariant decomposition problem of
the nucleon spin, it is instructive to compare a noticeable
difference between various
definitions of the physical component of the gauge field \cite{HJZ14}.
Namely, we compare
the LC gauge motivated definition, the temporal-gauge motivated one,
the spatial axial-gauge motivated one, and the Coulomb-gauge motivated one
given as 
\begin{eqnarray}
 A^k_{phys} \ &=& \ \frac{1}{D^+} \,F^{+k}, \label{Eq:LC_motivated} \\
 A^k_{phys} \ &=& \ \frac{1}{D^0} \,F^{0k}, \label{Eq:temporal_motivated} \\
 A^k_{phys} \ &=& \ \frac{1}{D^3} \,F^{3k}, \label{Eq:spatial_motivated}
\end{eqnarray}
with $k = 1, \ \mbox{or} \ 2$, and $D^+ \equiv \partial^+ - \,i \,g \,A^+$,
$D^0 \equiv \partial^0 - \,i \,g \,A^0$, $D^3 \equiv \partial^3 - \,i \,g \,A^3$,
while
\begin{equation}
 \bm{A}_{phys} \ = \ \bm{A} \ - \ \nabla \,\,\frac{1}{\nabla^2} \,\,
 \nabla \cdot \bm{A}. \label{Eq:Coulomb_motivated}
\end{equation}

In principle, all these definitions of the physical component of the gluon
give the corresponding definitions of the gluon spin by the relation,
\begin{equation}
 \Delta G = \frac{1}{2 \,P \cdot \tilde{n}} \,\langle P S \,|\,2 \,\mbox{Tr} 
 \left[ \epsilon^{jk}_\perp \,F^{j \tilde{n}} (0) \,A^k_{phys} (0) \right] \,
 |\, P S \rangle .
\end{equation}
Here, $\epsilon^{jk}_\perp \,(j, k = 1,2)$ is the antisymmetric tensor in the
transverse plane with $\epsilon^{12}_\perp = + \,1$. On the other hand,
$F^{i \tilde{n}} \equiv F^{i \lambda} \,\tilde{n}_\lambda$, where
$\tilde{n}^\mu = (1,0,0,-1) / \sqrt{2}$
for the choice (\ref{Eq:LC_motivated}), $\tilde{n}^\mu = (0,0,0,1)$ for the choice
(\ref{Eq:spatial_motivated}), and $\tilde{n}^\mu = (1,0,0,0)$ for the choices
(\ref{Eq:temporal_motivated}) and (\ref{Eq:Coulomb_motivated}).
Note however that a prominent feature of the LC gauge motivated choice of
the physical component is
that it is invariant under the Lorentz-boost along the 3-direction, i.e.
the direction of nucleon momentum,
which we have seen is a necessary condition for the definition of the gluon
spin corresponding to the DIS measurements. In fact, under the Lorentz boost
along the 3-direction given by (\ref{Eq:Boost}),
we can easily verify that the physical component in the LC-gauge motivated
definition is just invariant : 
\begin{eqnarray}
 A^k_{phys} \ &\equiv& \  
 \frac{1}{D^+} \,\,F^{+ k} \ = \  
 \frac{1}{\partial^+ - i \,g \,A^+} \,\,F^{+ k} \nonumber \\
 \ &\rightarrow& \ \frac{1}{\gamma \,(1-v) \,
 (\partial^+ - i \,g \,A^+)} \,\,\gamma \,(1-v) \, F^{+ k} \ \ \ \nonumber \\
 &=& \ \frac{1}{D^+} \,\,F^{+ k} \ = \ A^k_{phys} .
\end{eqnarray}
On the contrary, any other definitions of $A^k_{phys}$ is not invariant under
the boost. For example, the physical component with the temporal-gauge-motivated
choice transforms as
\begin{eqnarray}
 A^k_{phys} &\equiv& \frac{1}{D^0} \,F^{0k} \ = \ 
 \frac{1}{\partial^0 - \,i \,g \,A^0} \,F^{0k} \nonumber \\
 &\rightarrow& \frac{1}{\gamma \,[\,(\partial^0 - v \,\partial^3)
 \, - \,i \,g \,(A^0 - v \,A^3) \,]} \,\,\gamma \,(F^{0k} - v \,F^{3k}) 
 \nonumber \\
 \ &=& \ \frac{1}{D^0 - v \,D^3} \,\,(F^{0k} - v \,F^{3k}) ,
\end{eqnarray}
so that it is not clearly boost-invariant.
We therefore conclude that what plays a key role in the uniqueness problem of
the gauge-invariant decomposition of the nucleon spin is the Lorentz-frame
independence, or more precisely the Lorentz-boost invariance
along the direction of parent nucleon. 
This means that, somewhat unexpectedly, the relativity plays more crucial
role than the gauge principle in the unique definition of measurable
longitudinal gluon spin and consequently in the gauge-invariant decomposition
problem of the longitudinal nucleon spin.

Still nontrivial observation is as follows. In the free field limit, we can
always set both of the scalar and longitudinal components to be zero :
\begin{equation}
 A^0 \ = \ A^3 \ = \ 0.
\end{equation}
In this case, all of the general axial-gauge motivated definitions
of the physical component of the gluon reduce to the same expression :
\begin{eqnarray}
 \mbox{\tt LC} \ &:& \
 A^k_{phys} = \frac{1}{D^+} \,F^{+ k}
 \ \rightarrow \ \frac{1}{\partial^+} \,
 \partial^+ \,A^k = A^k , \\
 \mbox{\tt temporal} \ &:& \ 
 A^k_{phys} = \frac{1}{D^0} \,F^{0 k} \ 
 \ \rightarrow \ \frac{1}{\partial^0} \,
 \partial^0 \,A^k \ = A^k , \\
 \mbox{\tt axial} \ &:& \ 
 A^k_{phys} = \frac{1}{D^3} \,F^{3 k} \ 
 \ \rightarrow \ \frac{1}{\partial^3} \,
 \partial^3 \,A^k \ = A^k . \ \ \ \ \ \ 
\end{eqnarray}

This indicates perturbative equivalence of these three. 
In fact, the 1-loop anomalous dimension of the gluon spin operators
was calculated in \cite{Waka13} and it turned out that they are all equal :
\begin{eqnarray}
 &\,& \langle P S \,|\, \epsilon^{jk}_\perp \,
 F^{j+} \,A^k_{phys} \,|\,
 P S \rangle_G \,|_{A^+ = 0} \nonumber \\
 &\,& \hspace{-3mm} \ = \ \left[\,
 1 + \frac{\alpha_S}{4 \,\pi} \cdot 
 \frac{\beta_0}{\varepsilon} \right] \,
 \langle P S \,|\, \epsilon^{jk}_\perp \,
 F^{j+} \,A^k_{phys} \,|\,
 P S \rangle_G^{tree} , \hspace{5mm} \\
 &\,& \langle P S \,|\, \epsilon^{jk}_\perp \,
 F^{j0} \,A^k_{phys} \,|\,
 P S \rangle_G \,|_{A^0 = 0} \nonumber \\
 &\,& \hspace{-3mm} \ = \ \left[\,
 1 + \frac{\alpha_S}{4 \,\pi} \cdot 
 \frac{\beta_0}{\varepsilon} \right] \,
 \langle P S \,|\, \epsilon^{jk}_\perp \,
 F^{j0} \,A^k_{phys} \,|\,
 P S \rangle_G^{tree} , \\
 &\,& \langle P S \,|\, \epsilon^{jk}_\perp \,
 F^{j3} \,A^k_{phys} \,|\,
 P S \rangle_G \,|_{A^3 = 0}  \nonumber \\
 &\,& \hspace{-3mm} \ = \ \left[\,
 1 + \frac{\alpha_S}{4 \,\pi} \cdot 
 \frac{\beta_0}{\varepsilon} \right] \,
 \langle P S \,|\, \epsilon^{jk}_\perp\,
 F^{j3} \,A^k_{phys} \,|\,
 P S \rangle_G^{tree} ,
\end{eqnarray} 
where $\beta_0 = 11 - 2 \,n_f \,/\,3$.
(We recall that, for the LC-gauge and temporal-gauge motivated choices,
not only the divergent part but also the finite one in the 1-loop
corrections to the gluon spin operator were calculated
in \cite{HJZ14} and shown to coincide including the finite part.
However, the spatial
axial-gauge motivated choice was not investigated by them, because they
thought that this choice does not have a straightforward gluon spin
interpretation. We shall later show that this case also has helicity interpretation.)

The analysis above also reveals the following fact. In the free field
case, since we can eventually set $A^0 = A^3 = 0$, the physical components
of the gluon field is obviously the two transverse components $A^1$ and $A^2$.
Under the presence of the quark-gluon interaction or the color-charged
sources for the gluons, the situation is more complicated. 
It is true even in this case that the independent dynamical degrees of
freedom, which should be quantized, are the two transverse degrees of freedom,
and the other two components, i.e. the scalar component $A^0$ and the
longitudinal component $A^3$ are the dependent fields, which can be
expressed in terms of other dynamical degrees of freedom, i.e.
the two transverse components of the gluon field and the quark field.
(In the covariant
treatment of gauge theories, the scalar and longitudinal components
are also quantized. In this case, however, we must work in the
Hilbert space with indefinite metric.)
Nonetheless, because of the constraint
of the Gauss law, we cannot set both of $A^0$ and $A^3$ to be zero
simultaneously. (One might say that they still contain nontrivial physics
at least in our problem of a strongly-coupled bound system
of quarks and gluons.)
This is thought to be the reason why the
above mentioned various definitions of the physical component of the
gluon lead to physically different definitions of the gluon spin
in a nonperturbative sense.  

We emphasize that, if the definition of the gluon spin operator corresponding
to DIS observations is uniquely fixed, it also means that there is only one
(or two) nucleon
spin decomposition.
The explicit form of the gluon spin operator can naturally be read out from
the standard expression of the longitudinally polarized gluon distribution
$\Delta g(x)$ given by Manohar \cite{Manohar91} : 
\begin{eqnarray}
 &\,& \!\!\!\! \Delta g(x) \nonumber \\
 &\,& = \ \frac{i}{4 \,\pi \,x \,P^+} \int d \xi^- \,
 e^{i \,x \,P^+ \,\xi^-} \nonumber \\
 &\,& \hspace{8mm} \times \ \langle P S \,|\,
 \tilde{F}^{+ \lambda}_a (0) \,{\cal L}_{a b} [0, \xi_-] \,
 F^{+}_{b,\lambda} (\xi^-) \,|\,P S \rangle \nonumber \\
 &\,& = \, \frac{i}{4 \,\pi \,x \,P^+} \int d \xi^- \,
 e^{i \,x \,P^+ \,\xi^-} \nonumber \\
 \times \ &\,& \!\!\!\!\!\!  \langle P S \,|\,2 \,\mbox{Tr}
 \left[ \tilde{F}^{+ \lambda} (0) \,{\cal L} [0, \xi_-] \,
 F^{+}{}_{\lambda} (\xi^-) \,{\cal L} [\xi^-,0] \right] |\,P S \rangle , \ \ \ \ \ \ 
\end{eqnarray}
where $a,b$ are color indices. If we assume the usual principle-value
prescription to handle the singularity in the distribution functions
with use of the relation \cite{KT99},
\begin{equation}
 \int \,dx \,\,P \,\,\frac{1}{x} \,e^{\,i \,\lambda \,x} \ = \ 
 i \,\pi \,\epsilon (\lambda) ,
\end{equation}
where $\epsilon (x)$ is a step function defined as $\epsilon (x) = + \,1$
for $x > 0$ and $\epsilon (x) = - \,1$ for $x < 0$,
the first moment of $\Delta g(x)$ becomes
\begin{eqnarray}
 \Delta G \, &=& \, \int \,dx \,\Delta g(x) \nonumber \\
 \ &=& \ \frac{1}{2 \,P^+} \,\left(- \,\frac{1}{2} \right) \,
 \int \,d \xi^- \,\,\epsilon (\xi^-) \hspace{35mm} \nonumber \\
 \times \ &\,& \!\!\!\!\!\!\!\!\!\! 
 \langle P S \,|\, 2 \,\mbox{Tr} \! \left[ \tilde{F}^{+ \lambda} (0) \,
 {\cal L} [0,\xi^-] \, F^{+}{}_\lambda (\xi^-) \,{\cal L} [\xi^-, 0] \right] 
 \! |\,P S \rangle .
 \ \ \ \ \  
\end{eqnarray}
This can also be expressed as
\begin{eqnarray}
 &\,& \Delta G  =  \frac{1}{2 \,P^+} \,
 \langle P S \,|\,2 \,\mbox{Tr} \,\left[ \epsilon^{jk}_\perp \,
 F^{j +} (0) \, A^k_{phys} (0) \,\right]
 |\,P S \rangle , \ \ \ \ \ 
\end{eqnarray}
with the definition of the physical component of the gluon field \cite{Hatta11} : 
\begin{eqnarray}
 A^k_{phys} (0) 
 &=& - \,\frac{1}{2} \,
 \int \,d \xi^- \,\epsilon (\xi^-) \nonumber \\
 &\,& \hspace{8mm} \times \ {\cal L} [0, \xi^-] \,F^{+k} (\xi^-) \,
 {\cal L} [\xi^-, 0] . \hspace{5mm}
\end{eqnarray}
Hatta showed \cite{Hatta11} that the above-defined physical component
and the pure-gauge component given by $A^k_{pure} \equiv A^k - A^k_{phys}$
legitimately satisfy the general conditions (\ref{Eq:Gcond1}),
(\ref{Eq:Gcond2}) and (\ref{Eq:Gcond3}). We also point out that
the above definition of the physical component is formally equivalent
to (\ref{Eq:LC_motivated}).  
Anyhow, the above expression is gauge-invariant as well as Lorentz-boost
invariant along the 3-direction. 
This especially means that we can work in arbitrary gauges, although
$A^k_{phys}$ can be reduced to a local form only in the light-cone gauge.

What remains to be answered is the original question about apparent
contradiction with the standard textbook statement, which
tells us that the total angular momentum of the photon cannot be
gauge-invariantly decomposed into its spin and orbital parts.
A key to resolve this dilemma is the existence of particular
direction in the DIS physics, which is nothing but the direction of nucleon
momentum.
(The importance of a preferred direction dictated by the experimental
conditions for the gauge-invariant decomposition has also been
pointed out in footnote of \cite{BJ99} and page 14 of \cite{Lorce13A},
although the similarity with the photon spin measurement was
not mentioned in these papers.)
To understand it, we just recall that we have already encountered a similar
situation in the decomposition problem of the total photon angular momentum.
In the clearly written papers \cite{VanEN94A,VanEN94B}, 
Van Enk and Nienhuis 
argued that the total angular momentum of free electromagnetic field can
certainly be decomposed into spin and orbital parts without
causing conflict with gauge-invariance.
This separation is based on the familiar transverse-longitudinal
decomposition of the photon field. They clearly recognized that
this separation is not Lorentz invariant. The Lorentz-invariance
is not essential in this problem, however. The reason is that the measurement
of the spin and OAM of the photon is carried out by making use of
the interaction between atoms and the paraxial laser beam of a photon
and that this measurement is performed in a fixed laboratory frame.
One now realizes that a common feature of the gluon spin measurement and
the photon spin measurement is the existence of a particular spatial
direction. This particular direction in the photon spin
and OAM measurement is nothing but the direction of paraxial laser beam,
whereas it is the direction of nucleon momentum in the case of gluon spin
measurements.
We think that the final remark of the paper by Van Enk and Nienhuis
is extremely enlightening to understand the physics behind and we
quote it here for pedagogical reason. 
``The conclusion is that both ``spin'' and ``orbital'' angular momentum of a
photon are well defined and separately measurable.
This concerns all three components. However, only the components along the
propagation direction can be measured by detecting the change in internal and
external angular momentum of an atom, respectively.''
Also interesting to point out here is the following fact.
In more general nonparaxial case, it was argued that there is no clear
separation of the total photon angular momentum into its spin and OAM
parts \cite{BA94}.
Although the separation is not still impossible, a peculiarity
is that both the spin and OAM parts is generally dependent of the
photon helicity \cite{BAOA10}. This appears to be consistent with the
complexity of the transverse spin decomposition of the nucleon spin
\cite{JXY12,Leader13,HTY12,HKM14}
as compared with the longitudinal decomposition : the latter is of our main
concern in the present paper. 

Now we are ready to make a clear summary statement on the apparent
contradiction between the two observations pointed out in Introduction.
On the one hand, we know the classical statement in the standard textbooks
of electrodynamics that the total angular momentum of a massless particle
cannot be gauge-invariantly decomposed into its spin and orbital parts.
As already pointed out in Introduction, this statement should be replaced
by more precise one as follows.
The total angular momentum of a massless particle cannot be
decomposed into its spin and orbital parts so as to meet both requirements
of gauge-invariance and the Lorentz-frame independence. 
On the other hand, we know that the spin and OAM of a photon can be
separately measured. We conclude that what rescues this conflict
is the existence of a particular spatial direction.
Although the idea of {\it transversality} is certainly Lorentz-frame-dependent
in general, the (gauge-invariant) transverse component of the photon
or (gauge-covariant) physical components of the gluon can
consistently be defined with respect to this particular spatial direction.
Still, one should not forget about an important difference between
the measurement of the photon spin and that of the gluon spin in the nucleon : 
that is the role of relativity. 
Different from the photon spin problem, the relativity plays an essential
role in the decomposition problem of the total gluon angular momentum. 
What uniquely fix the nucleon spin decomposition corresponding to DIS
measurements is the requirement of boost-invariance along the direction
of the nucleon momentum. After all, the fact is that
the Lorentz symmetry plays more crucial role than the gauge symmetry
in the proper definition of the nucleon spin decomposition.

Although we believe that the argument above essentially demystifies the
controversial points in the gauge-invariant decomposition problem of the
nucleon spin, we think it instructive to inspect the physical contents of
the resultant gluon spin operator in some more detail.
In the LC gauge, the theoretical expression of the gluon spin reduces to
the following form : 
\begin{eqnarray}
 \Delta G = \frac{1}{2 \,P^+} \,\,
 \langle P S \,|\,(\bm{E}_\perp \times \bm{A}_\perp)^3
 \, + \, \bm{B}_\perp \cdot 
 \bm{A}_\perp \,|\, P S \rangle , \ \ \ \ \ 
 \label{Eq:GSpin}
\end{eqnarray}
where we have omitted the color indices, for brevity.
In the above equation, $\bm{A}_\perp$ should be understood to represent
the physical component in the light-cone gauge. 
We emphasize that the presence of the 2nd term is very important, because
the 1st term alone is not invariant under the boost along the 3-direction.
This can easily be verified from the following transformation properties
of the relevant quantities under the boost along the 3-direction.
\begin{eqnarray}
 E^1 \ \rightarrow \ \gamma \,(E^1 \ - \ v \,B^2), \ \ 
 E^2 \ \rightarrow \ \gamma \,(B^2 \ + \ v \,B^1), \ \ \ \\
 B^1 \ \rightarrow \ \gamma \,(B^1 \ + \ v \,E^2), \ \ 
 B^2 \ \rightarrow \ \gamma \,(B^2 \ - \ v \,E^1), \ \ \ 
\end{eqnarray}
and
\begin{equation}
 A^{1,2} \ \rightarrow \ A^{1,2} .
\end{equation}
We stress again that the boost-invariance of the physical or transverse
component $\bm{A}_\perp$ is guaranteed only for the light-cone gauge
or light-cone gauge motivated choice. 
Jaffe once tried to estimate the contributions of both terms of (\ref{Eq:GSpin})
in the bag model as well as in the quark model \cite{Jaffe96}.
Jaffe already recognized the fact that, since the sum is boost-invariant, the
above $\Delta G$ can in principle be calculated in any Lorentz-frame including
the rest frame of the nucleon, provided that the above $\bm{A}_\perp$ is the
gauge potential in the LC gauge.

What is curious here is the physical meaning of the unfamiliar 2nd term of
(\ref{Eq:GSpin}). Interestingly, it resembles the following quantity : 
\begin{eqnarray}
 S \ = \ \int \,\bm{B} \cdot \bm{A} \,\,d^3 x ,
\end{eqnarray}
except the absence of the 3-component in
$\bm{B}_\perp \cdot \bm{A}_\perp$.                   
In the field of space and laboratory plasma physics \cite{Berger99},
the above $S$ is called the magnetic helicity, and it is known to be
a topological invariant of magnetic field configuration. 
This might indicates that, if a topological configuration of the gluon field play some
role in the gluon spin, it is through this 2nd term. 

Aside from such a speculation, a perturbative consideration manifests
transparent physical meaning of the term $\bm{B}_\perp \cdot \bm{A}_\perp$.
Using the familiar free field expansion of the gauge potential,
\begin{eqnarray}
 &\,& \bm{A}_\perp (\bm{x},t) =  
 \int \! d^3 \tilde{k} \,
 \sum_{\lambda = \pm \,1} \nonumber \\ 
 &\,& \times \ [\,a (\bm{k},\lambda) \,
 \bm{\varepsilon} (\bm{k},\lambda) \,
 e^{- \,i \,k \cdot x} +  
 a^\dagger (\bm{k},\lambda) \,
 \bm{\varepsilon}^* (\bm{k},\lambda) \,
 e^{\,i \,k \cdot x} \,] , \hspace{8mm}
\end{eqnarray}
with $d^3 \tilde{k} = d^3 k \,/\,(2 \,\pi)^3 \,|\bm{k}|$, and with 
$\lambda$ representing the two helicity states of the gluon, one can
easily verify the following two relations : 
\begin{eqnarray}
 \int \,\bm{E}_\perp \times \bm{A}_\perp \,d^3 x
 = \int \,d^3 \tilde{k} \,\sum_{\lambda = \pm \,1} 
 \hat{\bm{k}}
 \,\,\lambda \,a^\dagger (\bm{k},\lambda) \,
 a (\bm{k},\lambda) , \ \ \ \ \ 
\end{eqnarray}
and
\begin{eqnarray}
 \int \,\bm{B}_\perp \cdot \bm{A}_\perp \,d^3 x
 = \int \,d^3 \tilde{k} \,\sum_{\lambda = \pm \,1} 
 \lambda \,a^\dagger (\bm{k},\lambda) \,a (\bm{k},\lambda) . \ \ \ \ 
\end{eqnarray}
Then, despite its unfamiliar appearance, the term
$\bm{B}_\perp \cdot \bm{A}_\perp$ also has the meaning of gluon
helicity at least in a perturbative sense.
Adding up the two terms, we thus find the following equation
\begin{eqnarray}
 &\,& \frac{1}{2} \,\int \,\left[\,
 (\bm{E}_\perp \times \bm{A}_\perp)^3 \, + \, 
 \bm{B}_\perp \cdot \bm{A}_\perp \,\right] \,d^3 x \ \ \ 
 \nonumber \\
 &\,& \hspace{6mm} = \ \sum_{\lambda = \pm \,1} \,
 \lambda \,\,
 a^\dagger (\bm{k},\lambda) \,a^(\bm{k},\lambda) .
\end{eqnarray}
As a consequence, we find that the sum of the two pieces in (\ref{Eq:GSpin})
reduces to the ordinary helicity operator of the gluon.

\section{Physical interpretation of the average transverse momentum
and longitudinal orbital angular momentum of quarks in the nucleon
defined by the Wigner distribution}
\label{sec:3}

The purpose of the present section to discuss the superiority or
inferiority of the canonical and mechanical decompositions of the nucleon
spin from the viewpoint of observation. 
Let us start with the definition of the average transverse momentum and
average longitudinal orbital angular momentum (OAM) of quarks in the
nucleon defined through the Wigner distribution as
\begin{eqnarray}
 \langle k^i_\perp \rangle^{\cal L} &=& \int \,dx \,\int \,d^2 b_\perp \,
 \int d^2 k_\perp \,\,k^i_\perp \,\rho^{\cal L} (x, \bm{b}_\perp, \bm{k}_\perp), 
 \ \ \ \ \label{Eq:Wigner_trans_mom} \\
 \langle L^3 \rangle^{\cal L} &=& \int \,dx \,\int \,d^2 b_\perp \,
 \int d^2 k_\perp \nonumber \\
 &\,& \hspace{18mm} \times \ (\bm{b}_\perp \times \bm{k}_\perp)^3 \,
 \rho^{\cal L} (x, \bm{b}_\perp, \bm{k}_\perp), \label{Eq:Wigner_OAM}
\end{eqnarray}
where $i = 1, \mbox{or} \ 2$. Here, the Wigner distribution is a 5-dimensional
phase-space distribution defined by
\begin{eqnarray}
 \rho^{\cal L} (x, \bm{b}_\perp, \bm{k}_\perp) &=& 
 \frac{1}{2} \,\int \frac{d^2 \bm{\Delta}_\perp}{(2 \,\pi)^2}
 \int \frac{d^2 \bm{\xi}_\perp \,d \xi^-}{(2 \,\pi)^3} \nonumber \\
 &\times& e^{\,- \,i \bm{\Delta}_\perp \cdot \bm{b}_\perp} \,
 e^{\,i \,(x \,P^+ \,\xi^- \,- \,\bm{k}_\perp \cdot \bm{\xi}_\perp)} \nonumber \\
 &\times& \langle p^\prime, s^\prime \,|\,
 \bar{\psi}(0) \,\gamma^+ \,
 {\cal L} [0,\xi] \,\psi (\xi) \,|\, p, s \rangle , \ \ \ \  
\end{eqnarray}
with $P = \frac{1}{2} (p^\prime + p)$ and $p^\prime - p = (0,\bm{\Delta}_\perp, 0)$. 
The Wigner distribution is generally dependent on the path of the
gage-link ${\cal L} [0,\xi]$ connecting the two space-time points $0$ and $\xi$.
The two physically interesting choices of the gauge-link paths are the
future-pointing light-like staple path ${\cal L}^{+LC}$ and the past-pointing
light-like staple path ${\cal L}^{-LC}$ defined as
\begin{eqnarray}
 &\,& {\cal L}^{\pm LC} [0, \xi] \equiv
 {\cal L}^{(st)} [0^- \,\bm{0}_\perp, \pm \infty^- \,\bm{0}_\perp] \nonumber \\
 &\,& \hspace{1mm} \times \ 
 {\cal L}^{(st)} [\pm \infty^- \,\bm{0}_\perp, \pm \infty^- \,\bm{\xi}_\perp] \,
 {\cal L}^{(st)} [\pm \infty^- \,\bm{\xi}_\perp, \xi^- \,\bm{\xi}_\perp] . 
 \ \ \ \ \ \  
\end{eqnarray}
Here, ${\cal L}^{(st)} [\xi, \eta]$
represent a straight-line path connecting directly the two space-time points
$\xi$ and $\eta$. (In the following, the suffix $(st)$ will be omitted for
brevity, when there is no possibility of misunderstanding.)
The two choices respectively corresponds to the kinematics of semi-inclusive
hadron productions and that of Drell-Yan processes.

For the average transverse momentum and average longitudinal OAM of
quarks respectively defined by (\ref{Eq:Wigner_trans_mom}) and
(\ref{Eq:Wigner_OAM}), Burkardt derived the following
relations \cite{Burkardt13}. First, for the average transverse momentum,
he showed that
\begin{equation}
 \langle k^i_\perp \rangle^{\pm LC} \ = \ \langle k^i_\perp \rangle_{mech}
 \ + \ \langle k^i_\perp \rangle^{\pm LC}_{int} . \label{Eq:Average_trans_mom}
\end{equation}
where
\begin{equation}
 \langle k^i_\perp \rangle_{mech} \ = \ \frac{1}{2 \,p^+} \,
 \langle p, s \,|\, \bar{\psi} (0) \,\frac{1}{i} \,D^i_\perp (0) \,\psi (0) \,|\, p,s \rangle .
\end{equation}
with $D^i_\perp \equiv \partial^i_\perp - i \,g \,A^i_\perp$, while
\begin{eqnarray}
 &\,& \langle k^i_\perp \rangle ^{\pm LC}_{int} = - \,\frac{1}{2 \,p^+} 
 \int_{0}^{\pm \infty} d \eta^- \,
 \langle p, s \,|\, \bar{\psi} (0) \,
 {\cal L} [0^- \,\bm{0}_\perp, \eta^- \,\bm{0}_\perp] \ \ 
 \nonumber \\
 &\,& \hspace{8mm} \times \ g \,F^{+i} (\eta^-,\bm{0}_\perp) \,
 {\cal L} [\eta^- \,\bm{0}_\perp, 0^- \,\bm{0}_\perp] \,\psi (0) \,|\, p,s \rangle. 
 \ \ \ \ \  
\end{eqnarray}
Similarly, for the average longitudinal OAM, he obtains
\begin{equation}
 \langle L^3 \rangle^{\pm LC} \ = \ \langle L^3 \rangle_{mech} \ + \ 
 \langle L^3 \rangle^{\pm LC}_{int},
\end{equation}
where
\begin{eqnarray}
 &\,& \langle L^3 \rangle_{mech} = {\cal N} \,\int \,d^2 r_\perp \,
 \epsilon^{ij}_\perp \,r^i_\perp \nonumber \\
 &\,& \hspace{1mm} \times \ \langle p, s \,|\, \bar{\psi} (0^-,\bm{r}_\perp) \,
 \gamma^+ \, \frac{1}{i} \,
 D^j_\perp (0^-, \bm{r}_\perp) \,\psi (0^-,\bm{r}_\perp) \,
 |\, p, s \rangle , \ \ \ \ \ \  
\end{eqnarray}
with ${\cal N} = 1 \,/\,(2 \,p^+ \int \,d^2 r_\perp)$, and
\begin{eqnarray}
 &\,& \langle L^3 \rangle^{\pm LC}_{int} \, = \, - \,{\cal N} \,\int \,d^2 r_\perp \,
 \int_{0}^{\pm \infty} \,d \eta^- \,\,\epsilon^{ij}_\perp \,r^i_\perp \nonumber \\
 &\,& \hspace{3mm} \times \ \langle p, s \,|\,\bar{\psi} (0^-,\bm{r}_\perp) \,
 \gamma^+ \, {\cal L} [0^- \,\bm{r}_\perp, \eta^- \,\bm{r}_\perp] \nonumber \\
 &\,& \hspace{3mm} \times \ g \,F^{+j} [\eta^-,\bm{r}_\perp]
 \,{\cal L} [\eta^- \,\bm{r}_\perp, 0^- \,\bm{r}_\perp] \,\psi (0^-,\bm{r}_\perp) \,
 |\, p,s \rangle . \label{Eq:Wigner_OAM_int} \ \ \ \ \ 
\end{eqnarray}
%
Burkardt gave the following physical interpretation to these equations.
For example, for the choice ${\cal L} = + LC$ corresponding to the
semi-inclusive DIS, he pays attention to the difference between
$\langle k^i_\perp \rangle^{+ LC}$ and $\langle k^i_\perp \rangle_{mech}$, i.e.
the 2nd term $\langle k^i_\perp \rangle^{+ LC}_{int}$
of (\ref{Eq:Average_trans_mom}).
According to him, this quantity can be interpreted as the change of
transverse momentum of the quark when it leaves the target after being
struck by the virtual photon in the semi-inclusive reactions.
The legitimacy of this interpretation can easily be convinced by taking
the light-cone gauge. In this gauge, the gauge-link along the light-cone
direction becoms unity and the relevant component of the field-strength
tensor reduces to
\begin{eqnarray}
 - \,\sqrt{2} \,g \,F^{+2} &=& - g \,F^{02} \ - \ g \,F^{32} \nonumber \\
 &=& g \, (E^2 \ - \ B^1) \ = \ g \,[\bm{E} \ + \  (\bm{v} \times \bm{B})]^2,
 \ \ \ \ \ \ \ 
\end{eqnarray}
which is nothing but the $y$-component of the color Lorentz force
acting on a particle that moves with the light velocity in the $- \,z$
direction.

Entirely analogously, the difference between $\langle L^3 \rangle^{+LC}$
and $\langle L^3 \rangle_{mech}$ can be interpreted as the change of
OAM of the struck quark by the torque of the color Lorentz force.

This is certainly a very natural physical interpretation.
Also interesting here is the relation with Hatta's finding \cite{Hatta12}.
Hatta showed that, because of the time-reversal and parity (PT) symmetry,
the average OAMs corresponding to the two paths ${\cal L} = \pm LC$
are exactly the same and that it coincides with the gauge-invariant (GI) canonical
OAM, i.e.
\begin{equation}
 \langle L^3 \rangle^{+ LC} \ = \ \langle L^3 \rangle^{- LC} \ = \ 
 \langle L^3 \rangle_{can}, 
\end{equation}
where
\begin{eqnarray}
 \langle L^3 \rangle_{can} &=& {\cal N} \,\int \,d^2 r_\perp 
 \,\epsilon^{ij}_\perp \,\,r^i_\perp \,
 \langle p, s \,|\, \bar{\psi} (0^-,\bm{r}_\perp) \,\gamma^+ \nonumber \\
 &\,& \hspace{5mm} \times \ 
 \frac{1}{i} \,D^j_{\perp,pure} (0^-, \bm{r}_\perp) \,
 \psi (0^-, \bm{r}_\perp) \,|\, p,s \rangle. \ \ \ \ \ \ 
\end{eqnarray}
with the definition of the pure-gauge covariant derivative,
$D^i_{\perp,pure} (\bm{r_\perp}) = \nabla_{r^i_\perp} - i \,g \,A^i_{pure} (\bm{r}_\perp)$.
Here, $A^\mu_{pure}$ is the pure-gauge component of the decomposition
$A^\mu = A^\mu_{phys} + A^\mu_{pure}$.
This means that the average longitudinal OAM defined through the Wigner
distribution coincide with the GI OAM not the mechanical one and that it is
process-independent. One might expect that a similar relation hold also
for the average transverse momentum, 
\begin{equation}
 \langle k^i_\perp \rangle^{\pm LC} \ \stackrel{?}{=} \ \langle k^i_\perp \rangle_{can},
\end{equation}
where the GI canonical transverse momentum is defined as
\begin{equation}
 \langle k^i_\perp \rangle_{can} \ = \ \frac{1}{2 \,p^+} \,
 \langle p, s \,|\, \bar{\psi} (0) \,\gamma^+ \,\frac{1}{i} \,
 D^i_{\perp,pure} (0) \,\psi (0) \,
 |\, p, s \rangle,
\end{equation}
In fact, Lorce claims in \cite{Lorce13B} that the momentum variable in
the Wigner distribution refers to the canonical momentum not the mechanical
momentum. In the following, we show that this statement is not always true
and then give a universally correct physical interpretation of the average
transverse momentum as well as the average longitudinal OAM defined
through the Wigner distribution.

To this end, we first recall the following fact.
According to Hatta \cite{Hatta12}, there are plural possibilities to define the
physical component of the gluon in the decomposition $A^\mu = A^\mu_{phys} + 
A^\mu_{pure}$. They are 
\begin{eqnarray}
 &\,& A^i_{phys} (0) \ \equiv \  - \,\int_{- \,\infty}^{+ \,\infty} \,d \eta^- \,\,
 (\pm \,\theta (\pm \,\eta^-)) \nonumber \\
 &\,& \hspace{1mm} \times \ {\cal L} [0^- \,\bm{0}_\perp, \eta^- \,\bm{0}_\perp]
 \,g \,F^{+i} (\eta^-, \bm{0}_\perp) \, 
 {\cal L} [\eta^- \,\bm{0}_\perp, 0^- \,\bm{0}_\perp] , \label{Eq:1st_choice} 
 \ \ \ \ \ \ \ 
\end{eqnarray}
or
\begin{eqnarray}
 &\,& A^i_{phys} (0) \ \equiv \ - \,\frac{1}{2} \,\int_{- \,\infty}^{+ \,\infty} \,d \eta^- \,\,
 \epsilon(\eta^-) \nonumber \\
 &\,& \hspace{1mm} \times \ {\cal L} [0^- \,\bm{0}_\perp, \eta^- \,\bm{0}_\perp]
 \,g \,F^{+i} (\eta^-, \bm{0}_\perp) \, 
 {\cal L} [\eta^- \,\bm{0}_\perp, 0^- \,\bm{0}_\perp] , \label{Eq:2nd_choice}
 \ \ \ \ \  
\end{eqnarray}
Here, $\theta (x)$ is the ordinary step function of Heaviside, while the sign
function $\epsilon (x)$ is defined as $1$ for $x > 0$ and $- \,1$ for $x < 0$.
As shown by Hatta \cite{Hatta12}, in the case of average longitudinal OAM,
any of the above choices for $A^i_{phys}$ gives the same answer
for $\langle L^3 \rangle^{\pm LC}$, which coincides with the canonical OAM of quarks.
This is due to the PT-even nature of the quantity $\langle L^3 \rangle$.
As one can convince easily, it is not necessarily true for the average transverse
momentum. If one adopts the first choice (\ref{Eq:1st_choice}), 
one certainly obtains
\begin{eqnarray}
 \langle k^i_\perp \rangle^{\pm LC} &=& \frac{1}{2 \,p^+} \,
 \langle p, s \,|\,\bar{\psi} (0) \,\gamma^+ \,\frac{1}{i} \,D^i_\perp (0) \,
 \psi (0) \,|\, p, s \rangle \nonumber \\
 &+& \frac{1}{2 \,p^+} \,\langle p, s \,|\,\bar{\psi} (0) \,\gamma^+ \,g \,
 A^i_{phys} (0) \,\psi (0) \,|\, p,s \rangle \nonumber \\
 &=& \frac{1}{2 \,p^+} \,\langle p, s \,|\,\bar{\psi} (0) \,\gamma^+ \,\frac{1}{i} \,
 D^i_{\perp, pure} (0) \,  \psi (0) \,|\, p, s \rangle \ \ \ \nonumber \\ 
 &\equiv& \ \langle k^i_\perp \rangle_{can},
\end{eqnarray}
which in fact coincides with the definition of the GI canonical transverse momentum
in conformity with the conclusion of Lorce \cite{Lorce13B}.
On the other hand, however, it is a well known fact that the average transverse
momentum corresponding to the future-pointing staple light-cone path and
the past-pointing stale light-cone path have different signs
as \cite{Collins02,BJY03,BMP03}
\begin{equation}
 \langle k^i_\perp \rangle^{- LC} \ = \ - \, \langle k^i_\perp \rangle^{+ LC}.
\end{equation}
Then, the canonical transverse momentum defined as above is not a universal
quantity, i.e. it is process-dependent.
(Here the word "path-dependence" and "process-dependence"
should not be taken too generically. What is meant here is
actually the dependence on the direction of the staple-like
gauge link, which respectively correspond to two physical
processes, i.e. the semi-inclusive hadron productions and
the Drell-Yan processes.)

More natural choice of $A^i_{phys}$ is therefore given by (\ref{Eq:2nd_choice}).
In this case, by using the mathematical identity
\begin{equation}
 \pm \,\theta (\pm \,\eta^-) \ = \ \frac{1}{2} \,[\,
 \epsilon (\eta^-) \ \pm \ 1] ,
\end{equation}
we obtain
\begin{eqnarray}
 &\,& \langle k^i_\perp \rangle^{\pm LC} \ = \ \frac{1}{2 \,p^+} \,
 \langle p, s \,|\,\bar{\psi} (0) \,\gamma^+ \,\frac{1}{i} \,D^i_\perp (0) \,\psi (0) \,
 |\, p, s \rangle
 \nonumber \\
 &\,& \hspace{15mm} + \ \ \frac{1}{2 \,p^+} \,
 \langle p, s \,|\,\bar{\psi} (0) \,\gamma^+ \,g \,A^i_{phys} (0) \,\psi (0) \,
 |\,p, s \rangle
 \nonumber \\
 &\,& \hspace{15mm} \mp \,\frac{1}{4 \,p^+} \,\int_{- \,\infty}^{+ \,\infty} \,d \eta^- \,
 \langle p, s \,|\, \bar{\psi} (0) \,\gamma^+ \,{\cal L} [0^-, \eta^-] 
 \ \ \ \ \ \nonumber \\
 &\,& \hspace{25mm} \times \ g \,F^{+i} (\eta^-) \,
{\cal L} [\eta^-, 0^-] \,\psi (0) \,\,|\,p, s \rangle , 
 \label{Eq:Average_trans_mom_3term} \ \ \ 
\end{eqnarray}
where $A^i_{phys}$ corresponds to the choice (\ref{Eq:2nd_choice}).
The 1st and the 2nd terms are respectively the mechanical transverse momentum
$\langle k^i_\perp \rangle_{mech}$ and the potential momentum
$\langle k^i_\perp \rangle_{pot}$. Combining these terms, the physical component
contained in $D^i_\perp$ cancels with the 2nd term so that the sum reduces to
the definition of the canonical transverse momentum $\langle k^i_\perp \rangle_{can}$.
Obviously, for this 2nd choice of $A^i_{phys}$, the definition of the canonical
momentum is path-independent or process-independent. However, we now have
\begin{equation}
 \langle k^i_\perp \rangle^{\pm LC} \ \neq \ \langle k^i_\perp \rangle_{can}.
\end{equation}
The argument above proves non-universal nature of the statement
in \cite{Lorce13B} that the momentum
variable in the Wigner distribution refers to the canonical momentum not
the mechanical momentum. In our opinion, the above-mentioned
arbitrariness in the definition of the canonical transverse momentum is an
indication of mathematical or theoretical (rather than physical) nature of the
canonical transverse momentum in contrast to mechanical transverse momentum
with more physical meaning. (Note that there is no ambiguity in the definition
of the latter. As we have repeatedly
emphasized \cite{Waka10,Waka11A,Waka12,Review_Waka14}, physical nature of the
mechanical momentum as compared with the canonical momentum is reflected
in the fact that the former not the latter appears in the equation of motion with
Lorentz force.)

Interestingly, if we adopt the 2nd option (\ref{Eq:2nd_choice}) for $A^i_{phys}$,
not only the mechanical transverse momentum but also the canonical transverse
momentum vanish due to the PT symmetry, 
\begin{equation}
 \langle k^i_\perp \rangle_{mech} \ = \ \langle k^i_\perp \rangle_{can} \ = \ 0.
\end{equation}
This means that only the last term of ({\ref{Eq:Average_trans_mom_3term})
contributes to the average transverse
momentum $\langle k^i_\perp \rangle^{\pm LC}$. As is well-known, this
3rd term can be related to the gluon-pole term of the twist-3 quark gluon
correlation function $\Psi_F (x, x^\prime)$ known as the 
Efremov-Teryaev-Qiu-Sterman (ETQS) 
function \cite{ET85,QS91,QS92}, which is defined by
\begin{eqnarray}
 &\,& \int \,\frac{d \xi^-}{2 \,\pi} \int \,\frac{d \eta^-}{2 \,\pi} \,
 e^{\,i \,p^+ \,\xi^- \,x} \,e^{\,i \,p^+ \,\eta^- \,(x^\prime - x)} \nonumber \\
 &\,& \times \,\langle p, s \,|\, \bar{\psi} (0) \,\gamma^+ \,
 {\cal L} [0^-,\eta^-] \,g \,F^{+i} (\eta^-) \,{\cal L} [\eta^-, \xi^-] \,
 \psi (\xi) \,|\, p, s \rangle \nonumber \\
 &\,& \hspace{1mm} = \ \frac{1}{p^+} \,\,\epsilon^{ij}_\perp \,s^j_\perp \,\,
 \Psi_F (x^\prime, x) \ + \ \cdots .
\end{eqnarray}
In fact, it is an easy exercise to show that
\begin{equation}
 \langle k^i_\perp \rangle^{\pm LC} \ = \ \frac{1}{2} \,\epsilon^{ij}_\perp \,
 s^j_\perp \,\,(\mp \,\pi) \,\,\int \,dx \,\Psi_F (x,x) . 
 \label{Eq:Average_trans_mom_ETQS}
\end{equation}
On the other hand, the average transverse momentum defined by the
Wigner distribution can also be expressed
with the transverse-momentum-dependent (TMD) distribution.
By starting with the relation
\begin{eqnarray}
 \langle k^i_\perp \rangle^{\pm LC} &=& \int \,dx \int \,d^2 b_\perp 
 \int \,d^2 k_\perp \,\,k^i_\perp \,\,
 \rho^{\gamma^+} (x, \bm{b}_\perp, \bm{k}_\perp) \nonumber \\
 &=& \int \,dx \int \,k^2_\perp \,\,\tilde{\rho}^{\gamma^+} 
 (x,\bm{k}_\perp, \bm{\Delta}_\perp)  |_{\Delta_\perp = 0} ,
\end{eqnarray}
together with the widely-used parametrization of the generalized
TMD (GTMD) \cite{MMS09},
\begin{eqnarray}
 &\,& \tilde{\rho}^{\gamma^+} (x,\bm{\Delta}_\perp,\bm{k}_\perp) \nonumber \\
 &\,& \hspace{2mm} = \  
 \frac{1}{2 \,p^+} \left[\,\gamma^+ \,F_{11} + 
 \frac{i \,\sigma^{i+} \Delta^i_\perp}{2 \,M_N} \,(2 \,F_{13} - F_{11}) \right.
 \nonumber \\
 &\,& \hspace{2mm} + \ \left. \frac{i \,\sigma^{i +} \,k^i_\perp}{2 \,M_N} \,\,2 \,F_{12}
 \ + \ \frac{i \,\sigma^{ij} \,k^i_\perp \,\Delta^j_\perp}{M_N^2} \,F_{14} \,\right] \,
 u (p,s) , \label{Eq:GTMD} \ \ \ \ \ \ 
\end{eqnarray}
one can show that
\begin{equation}
 \langle k^i_\perp \rangle^{\pm LC} \ = \ - \,\frac{1}{2} \,\,\epsilon^{ij}_\perp \,
 s^j_\perp \,\,\int \,d x \int \,d^2 k_\perp \,\frac{\bm{k}_\perp^2}{M_N} \,
 f_{1T}^\perp (x,\bm{k}_\perp^2). \label{Eq:Average_trans_mom_Sivers}
\end{equation}
Here, $f_{1T}^\perp$ is the familiar Sivers function \cite{Sivers90,Sivers91},
related to the imaginary part of the GTMD $F_{12}$ as
\begin{equation}
 f_{1T}^\perp (x,\bm{k}^2_\perp) \ = \ \mbox{Im} \,
 F_{12} \,(x, \xi = 0, \bm{k}_\perp^2, \bm{k}_\perp \cdot \bm{\Delta}_\perp = 0,
 \bm{\Delta}^2 = 0).
\end{equation}
Comparing (\ref{Eq:Average_trans_mom_ETQS}) and 
(\ref{Eq:Average_trans_mom_Sivers}), we therefore reproduce the
well-known relation between the Sivers function and the ETQS function given
as \cite{BMP03}
\begin{equation}
 \int \,d^2 k_\perp \,\frac{\bm{k}_\perp^2}{M_N} \,f_{1T}^\perp (x,\bm{k}_\perp^2)
 \ = \ \mp \,\pi \,\,\Psi_F (x,x) . \label{Eq:Sivers_ETQS}
\end{equation}

In the above analysis, we have shown that the identification of the average
transverse momentum of quarks defined through the Wigner distribution
with the canonical momentum is not necessarily justified and that the definition
of the canonical transverse momentum has intrinsic ambiguity. What is a
universally correct physical interpretation of the Wigner-distribution-based
average transverse momentum, then ? 
The answer can easily be read out from the paper by Burkardt \cite{Burkardt13}.
Taking the semi-inclusive DIS case as a concrete example, one is
allowed to say that the average
transverse momentum of quarks defined by the Wigner distribution represents
the asymptotic momentum of a quark after it leaves the target.
Note that this interpretation holds independently of the definitions of
the canonical transverse momentum. Naturally, how to relate this asymptotic
momentum of quarks to observables is a highly nontrivial question, because
of the color confinement of QCD, which does not allow the existence of
free quarks. Nevertheless, to grasp the physical meaning of the average
transverse momentum defined by the Wigner distribution,
it may be instructive to imagine a very hard quark jet produced
in the above-mentioned semi-inclusive DIS. The produced parent quark
is supposed to be fragmented into several hadrons running fast.
If one can measure the transverse momenta of all these hadrons,
one can in principle reconstruct the transverse
momentum of the original quark. Needless to say, the fact is not so
simple, because the fragmentation process occurs through the interaction
with the residual spectator. The detail may not be unrelated to the definition
of the quark jet algorithm. 
Nonetheless, we believe that the above gedankenexperiment clarifies
the physical interpretation of the average transverse momentum defined
through the Wigner distribution. Remember that this quark transverse momentum
in the asymptotic region is related to the Sivers function on the one hand, 
and the gluon-pole term of the ETQS function on the other hand.

Next, we shall discuss the interpretation of the average longitudinal
OAM defined through the Wigner distribution, by paying attention to the
similarity and the difference with the case of average transverse momentum.
We first point out that, using the technique as described
in \cite{HMS04}, the final- or initial-state interaction term
given by (\ref{Eq:Wigner_OAM_int}) can be written as
\begin{eqnarray}
 &\,& \langle L^3 \rangle^{\pm LC}_{int} \ = \ - \,{\cal N} \,d^2 r_\perp \,\,
 \epsilon^{ij}_\perp \,\,\left( - \,i \,\frac{\partial}{\partial \Delta^i_\perp} \right) \,
 \int_0^{\pm \infty} \,d \eta^- \nonumber \\
 &\,& \hspace{6mm} \times \ \langle p, \bm{\Delta}_\perp \,/\,2, s \,|\,
 \bar{\psi} (0^-,\bm{r}_\perp) \,\gamma^+  \nonumber \\
 &\,& \hspace{6mm} \times \ {\cal L} [0^- \,\bm{r}_\perp, \eta^- \,\bm{r}_\perp] \,
 g \,F^{+j} (\eta^-, \bm{r}_\perp) \,
 {\cal L} [\eta^- \,\bm{r}_\perp, 0^- \,\bm{r}_\perp] \nonumber \\
 &\,& \hspace{6mm} \times \ \psi (0^-, \bm{r}_\perp) \,| \,
 p^+, - \,\bm{\Delta}_\perp \,/\,2, s \rangle |_{\bm{\Delta}_\perp = 0} . \ \ \ \ \ 
\end{eqnarray}
With use of the identity
\begin{eqnarray}
 &\,& \int_0^{\pm \,\infty} \,d^- \eta \,f (\eta^-) \ = \ 
 \int_{- \,\infty}^{+ \,\infty} \,d \eta^- \,\left( \pm \,\theta(\pm \,\eta^-) \right) \,
 f(\eta^-) \nonumber \\
 &\,& = \ \int_{- \,\infty}^{+ \,\infty} \,d \eta^- \,\left( - \,\frac{1}{2 \,\pi} \right) \,
 \int_{- \,\infty}^{+ \,\infty} \,d x^\prime \,\frac{i}{x^\prime - x \mp i \,\varepsilon} 
 \nonumber \\
 &\,& \hspace{33mm} \times \ e^{\,i \,p^+ \,\eta^- \,(x^\prime - x)} \,f (\eta^-) ,
\end{eqnarray}
it can further be transformed into
\begin{eqnarray}
 &\,& \langle L^3 \rangle^{\pm LC} =  
 {\cal N} \,p^+ \int d^2 r_\perp \,
 \epsilon^{ij}_\perp \,
 \left( - \,i \,\frac{\partial}{\partial \Delta^i_\perp} \right) 
 \int \,\frac{d \xi^- \,d \eta^-}{(2 \,\pi)^2} 
 \nonumber \\
 &\,& \times \ \int \,d x \int \,d x^\prime \,
 \frac{i}{x^\prime - x \mp \, i \,\varepsilon} \,
 e^{\,i \,x \,p^+ \,\xi^-} \,\,e^{\,i \,p^+ \,\eta^- \,(x^\prime - x)} \nonumber \\
 &\,& \hspace{6mm} \times \ \langle p^+, \bm{\Delta}_\perp / 2, s^\prime \,| \,
 \bar{\psi} (0^-, \bm{r}_\perp) \,
 \gamma^+  \,{\cal L} [0^- \,\bm{r}_\perp, \eta^- \,\bm{r}_\perp] \nonumber \\
 &\,& \hspace{6mm} \times \ g \,F^{+j} (\eta^-, \bm{r}_\perp) \,
 \, {\cal L} [\eta^- \,\bm{r}_\perp, 0^- \,\bm{r}_\perp] \nonumber \\
 &\,& \hspace{6mm} \times \ \psi (0^-, \bm{r}_\perp) \,| \,
 p^+, - \,\bm{\Delta}_\perp / 2, s \rangle |_{\bm{\Delta}_\perp = 0} . 
 \label{Eq:OAM_int}  
\end{eqnarray}
As shown by Hatta and Yoshida \cite{HY12}, the quark-gluon correlation function
appearing in the above equation can be parametrized as
\begin{eqnarray}
 &\,& (p^+)^2 \,\int \,\frac{d \xi^-}{2 \,\pi} \int \,\frac{d \eta^-}{2 \,\pi} \,\,
 e^{\,i \,p^+ \,\xi^- \,x} \,\,e^{\,i \,p^+ \,\eta^- \,(x^\prime - x)} \nonumber \\
 &\,& \times \ \langle p^+, \bm{\Delta}_\perp / 2, \,s^\prime \,|\,
 \bar{\psi} (0^-,\bm{r}_\perp) \,\gamma^+ \nonumber \\
 &\,& \times \ {\cal L} [0^- \,\bm{r}_\perp, \eta^- \,\bm{r}_\perp] \,
 g \,F^{+j} (\eta^-, \bm{r}_\perp) \,
 {\cal L} [\eta^- \,\bm{r}_\perp, 0^- \,\bm{r}_\perp] \nonumber \\
 &\,& \times \ \psi (0^-, \bm{r}_\perp) \,| \,
 p^+, - \,\bm{\Delta}_\perp / 2, s \rangle \nonumber \\
 &\,& = \ \epsilon^{ij}_\perp \,p^+ \,\bar{s}^j_\perp \,
 \Psi_F (x,x^\prime) + \epsilon^{ij}_\perp \,\bar{s}^+ \,\Delta^j_\perp \,
 \Phi_F (x,x^\prime) + \cdots , \ \ \ \ \ 
\end{eqnarray}
with $\bar{s}^\mu = (s^{\prime \mu} + s^\mu) \,/\,2$. (Our normalization of
the nucleon spin vector is that $s^2 = - \,1$.)
Here, $\Psi_F (x,x^\prime)$ is the familiar ETQS function, whereas $\Phi_F (x,x^\prime)$
is a totally new quark-gluon correlation function that does not appear in the
forward limit $\bm{\Delta}_\perp \rightarrow 0$.
They have the following symmetry properties
\begin{equation}
 \Psi_F (x, x^\prime) \ = \ \Psi_F (x^\prime,x), \ \ 
 \Phi_F (x,x^\prime) \ = \ - \,\Phi_F (x^\prime, x) . \label{Eq:symmetry} \ \ 
\end{equation}
Inserting the above parametrization (\ref{Eq:F-type_correlation}) into
(\ref{Eq:OAM_int}), 
the $\Psi_F (x,x^\prime)$ term drops out and one obtains 
\begin{equation}
 \langle L^3 \rangle^{\pm LC}_{int} \ = \ s^+ \,\,\int \,d x \int \,d x^\prime \,\,
 \frac{1}{x^\prime - x \mp \,i \varepsilon} \,\Phi_F (x,x^\prime) .
 \label{Eq:F-type_correlation}
\end{equation}
Next, because of the antisymmetric property (\ref{Eq:symmetry}) of 
the function $\Phi_F (x,x^\prime)$, the gluon-pole term does not
contribute to the above integral, and only the principle part remains, thereby
leading to
\begin{eqnarray}
 \langle L^3 \rangle^{\pm LC}_{int} &=& s^+ \,\,\int \,d x \int \,d x^\prime \,\,
 {\cal P} \,\frac{1}{x^\prime - x} \,\,\Phi_F (x, x^\prime) \nonumber \\
 &=& \ \langle L^3 \rangle_{pot}.
\end{eqnarray}
This last formula is just the expression of the potential angular
momentum $\langle L^3 \rangle_{pot}$ given by Hatta \cite{Hatta12} except
for the overall sign difference from our definition.
(The reason of our definition is explained
in our review paper \cite{Review_Waka14}.) We therefore reconfirms that
\begin{equation}
 \langle L^3 \rangle^{\pm LC} \ = \ \langle L^3 \rangle_{mech} \ + \ 
 \langle L^3 \rangle_{pot} \ = \ \langle L^3 \rangle_{can},
\end{equation}
which shows that the average longitudinal OAM of quarks defined through the
Wigner distribution reduces to the GI canonical OAM.
Undoubtedly, the process-independence of this canonical OAM is a
consequence of vanishing gluon-pole contribution, which is totally
different from the average transverse momentum case.
Alternatively, it can be traced back to
the PT even nature of the longitudinal OAM. What is interesting to remember
here is the universal nature of our interpretation of the physical quantities
defined through the Wigner distribution, which holds commonly for the
average transverse momentum and longitudinal OAM.
We already gave an interpretation that the average transverse momentum
defined by the Wigner distribution gives the momentum of the quark in the
asymptotic region after leaving the spectator in the semi-inclusive DIS
processes. Note that exactly the same interpretation holds also for the
average longitudinal OAM. Namely, we can interpret that the average
longitudinal OAM defined by the Wigner distribution represents the
OAM of the quark in the asymptotic region after leaving the spectator. 
If it were not for the color confinement, this quark would be free.
This naturally explains the reason why the canonical OAM (basically
free field OAM) not the mechanical OAM appears in the
Wigner-function-motivated definition.
As we shall discuss shortly, however, this does not mean that the
canonical OAM is easier to
measure within the standard theoretical framework of DIS scatterings.

At this stage, it is useful to recall the fact that, as first pointed out by
Lorce and Pasquini \cite{LP11},
the average OAM defined by the Wigner distribution
can also be expressed with the GTMD $F_{14}$ appearing in the parametrization
(\ref{Eq:GTMD}). The relation is given by
\begin{eqnarray}
 &\,& \langle L^3 \rangle^{\pm LC} \ = \ \langle L^3 \rangle_{can} \nonumber \\
 &\,& \hspace{3mm} = \ s^+ \,\biggl\{ - \int dx \int d^2 k_\perp \,\,
 \frac{\bm{k}_\perp^2}{M_N^2}  \nonumber \\
 &\,& \hspace{8mm} \times \ F_{14} (x,\xi = 0, \bm{k}_\perp^2, 
 \bm{k}_\perp \cdot \bm{\Delta}_\perp = 0,
 \,\bm{\Delta}_\perp^2 = 0) \,\biggr\} . \hspace{8mm} 
\end{eqnarray}
Since this should coincides with the sum of $\langle L^3 \rangle_{mech}$
and $\langle L^3 \rangle_{pot}$ and since we know that the mechanical
OAM $\langle L^3 \rangle_{mech}$ can be related to the known GPD
$G_2 (x,\xi,t)$ as \cite{PPSS00,KP04,HMS04,HY12}
\begin{equation}
 \langle L^3 \rangle_{mech} \, = \, s^+ \,\left\{\, - \int dx \,x \,
 G_2 (x, \xi = 0, \bm{\Delta}_\perp^2 = 0) \,\right\}, \label{Eq:G2_SR}
\end{equation}
we conclude that the following relation must hold
\begin{eqnarray}
 &-& \int \,dx \int \,d^2 k_\perp \,\frac{\bm{k}_\perp^2}{M_N^2} 
 \,F_{14} (x, 0, 0, 0, 0) \ \ \ \ \ \ \ \nonumber \\
 &\,& \ = \ - \, \int \,dx \,\,x \,G_2 (x, 0, 0) \nonumber \\
 &\,& \ + \ 
 \int \,dx \int \,d x^\prime \,{\cal P} \,\,
 \frac{1}{x^\prime - x} \,\Phi_F (x,x^\prime) . \label{Eq:F14_QG_correlation}
\end{eqnarray}
This corresponds to the relation (\ref{Eq:Sivers_ETQS}) between the Sivers
function and the ETQS function obtained from the analysis of the average
transverse momentum of quarks. 
We emphasize the fact that the quark-gluon correlation function appearing
in the above equation has nothing to do with the familiar ETQS function.
This in turn implies non-quantitative nature of the frequently-claimed
connection between the Sivers function and the quark orbital angular
momentum.
From the practical viewpoint, the above relation
(\ref{Eq:F14_QG_correlation}) may not be a very useful
relation. To see it, we shall first concentrate on the l.h.s. of the above
relation. The problem is that, as first emphasized by Courtoy et al., the
GTMD $F_{14}$ drops out in both the TMD and GPD
formulation \cite{CGHLR14A,CGHLR14B}.
The TMD factorization and/or the collinear factorization are the bases of
model-independent extraction of TMDs, PDFs, and GPDs. 
Then, if it does not appear in either factorization, there is no way to extract
it in a model-independent way.
We recall that the situation is very similar to the $D$-state probability of
the deuteron. It is generally believed that the OAMs of a constituent in a
composite system is not a direct observable. This is easily convinced
by considering the decomposition of the total spin of the deuteron, which is
the simplest nucleus aside from the proton itself. Using the standard-form
wave function of the deuteron given as
\begin{equation}
 \psi_d (\bm{r}) \ = \ \frac{1}{\sqrt{4 \,\pi}} \,
 \left[\, u(r) \ + \ \frac{S_{12} (\hat{\bm{r}})}{\sqrt{8}} \,\,
 w(r) \,\right] \,\chi_{S = 1} ,
\end{equation}
where $u(r)$ and $w(r)$ are the radial wave functions of the $S$- and $D$-states,
while $\chi_{S=1}$ is the spin state of the deuteron, the total angular momentum
of the deuteron can be decomposed into the orbital and intrinsic spin parts as
\begin{eqnarray}
 1 \ = \ \langle J^3 \rangle &=& 
 \langle (\hat{\bm{r}} \times \hat{\bm{p}})^3 \rangle
 \ + \ 
 \langle \hat{S}^3 \rangle \nonumber \\
 &=& \hspace{4mm}
 \frac{3}{2} \,P_D \ + \ \left(\,P_S \ - \ \frac{1}{2} \,P_D \,\right) , 
\end{eqnarray}
where $P_S$ and $P_D$ are the $S$- and $D$-state probabilities defined by
\begin{equation}
 P_S \ = \ \int_0^\infty \,[u(r)]^2 \,r^2 \,d r, \ \ 
 P_D \ = \ \int_0^\infty \,[w(r)]^2 \,r^2 \,d r .
\end{equation}
One sees that the OAM contribution to the deuteron spin is $3/2$ times the
$D$-state probability. However, it is a well-known fact that the $D$-state
probability of the deuteron is not a direct observable \cite{Amado79,Friar79}.
The point is that bound-state wave functions, especially its short range part,
are not direct observables. In fact, it was explicitly demonstrated
in \cite{BFRS07} that the
$D$-state probability of the deuteron can drastically be altered by varying
the short-distance part of the nuclear force and consequently the nuclear
wave function without affecting the direct observables like the deuteron
binding energy or the asymptotic $D/S$ ratio, etc.
To be more general, the nuclear wave functions of any nucleus and
consequently the momentum distribution of the nucleon in the nucleus are
not direct observables.   
The situation is very different for the momentum distributions of
partons in the nucleon. Owing to the factorization theorem, the momentum
distributions of partons in the nucleon are believed to be model-independently
extracted, so that they are thought to be observables or at least
quasi-observables.

Coming back to the $F_{14}$ sum rule, the GTMD $F_{14}$ appears in neither
of the TMD factorization scheme nor the collinear (GPD) factorization
scheme, so that there is no way to extract it model-independently.
Needless to say, although the GTMD $F_{14}$ would not correspond to a
direct observable, it does not mean that it is useless.
From the theoretical viewpoint, $F_{14}$
can be calculated in some models or within the framework of lattice QCD,
so that it does give an useful insight into the spin contents of the nucleon.
The situation is again very similar to the $D$-state probability of
the deuteron, which gives valuable information on the spin
structure of the deuteron, even though it is a theoretical-scheme
dependent quantity.

Let us next turn to the discussion on the r.h.s. of the relation (41).
To clearly understand its significance from the observational viewpoint,
we think it instructive to recall the
key factor in the derivation of the important relation
(\ref{Eq:G2_SR}), which was first discovered by
Polyakov et al. \cite{PPSS00,KP04} and later
reconfirmed by several authors \cite{HMS04,HY12}.
The starting point is the identity
\begin{equation}
 0 \ = \langle \bar{\psi} (0) \,\gamma^i \!\not\!n \,{\cal L} [0,\lambda] 
 \not\!\!D (\lambda) \,\psi (\lambda) \rangle ,
\end{equation}
with $\langle \cdots \rangle = \langle p^\prime, s^\prime \,|\,\cdots \,|\,
p,s \rangle$, which holds exactly owing to the QCD equation of motion
$\not\!\!D (\lambda) \,\psi (\lambda) = 0$ \cite{Ratcliffe86,BB91}.
Using the so-called Chisholm identity
\begin{eqnarray}
 \gamma^\mu \,\gamma^\nu \,\gamma^\lambda \ &=& \  
 g^{\mu \nu} \,\gamma^\lambda \ + \ g^{\nu \lambda} \,\gamma^\mu
 \nonumber \\
 &-& \ g^{\mu \lambda} \,\gamma^\nu \ + \ i \,
 \epsilon^{\mu \nu \lambda \rho} \,\gamma_\rho \, \gamma_5 ,
\end{eqnarray}
one thus obtains
\begin{eqnarray}
 0 \ &=& \ i \,\frac{\partial}{\partial \lambda} \,\langle \bar{\psi} (0) \,
 \gamma^i \,{\cal L} [0,\lambda] \,\psi (\lambda) \rangle \nonumber \\
 &-& \ \langle \bar{\psi} (0) \,\gamma^+ \,{\cal L} [0,\lambda] \,
 D^i (\lambda) \,\psi (\lambda) \rangle \nonumber \\
 &+& \ i \,\epsilon^{i + \beta \rho} \,\langle \bar{\psi} (0) \,
 \gamma_\rho \,\gamma_5 \,{\cal L} [0,\lambda] \,D_\beta (\lambda) \,
 \psi (\lambda) \rangle . \label{Eq:QCD_identity}
\end{eqnarray}
Next, we take the Fourier transform over the variable $\lambda$.
Based on the parametrization of the twist-2 and
twist-3 GPDs \cite{PPSS00,KP04,HY12}, 
\begin{eqnarray}
 &\,& \int \,\frac{d \lambda}{2 \,\pi} \,
 e^{\,i \,\lambda \,x} \,
 \langle \bar{\psi} (0) \,
 \gamma^i \,{\cal L} (0, \lambda) \,\psi (\lambda) \rangle
 \nonumber \\
 &\,& = \ \bar{u} (p^\prime, s^\prime) \,
 \left[\,\frac{\Delta^i_\perp}{2 \,M_N} \,\,G^q_1 \ + \ \gamma^i \,
 (\,H^q + E^q + G^q_2 \,) \right. \ \ \ \ \ \nonumber \\
 &\,& \hspace{2mm} + \ \left. \frac{\Delta^i_\perp \,\gamma^+}{P^+} \,\,G^q_3
 \ + \ \frac{i \,\epsilon^{ij}_T \,\Delta^j_\perp \,\gamma^+\,\gamma_5}{P^+} \,
 \,G^q_4 \,\right] \,u (p, s) , \ \ \  
\end{eqnarray}
together with the identity
\begin{equation}
 \bar{u} (p^\prime,s^\prime) \,\gamma^i \,u (p, s) \ = \ 
 - \,\epsilon^{ij}_\perp \,\Delta^j_\perp \,s^+ ,
\end{equation}
the 1st term on the r.h.s. of (\ref{Eq:QCD_identity}) can be expressed as
\begin{eqnarray}
 &\,& \int \,\frac{d \lambda}{2 \,\pi} \,e^{i \,\lambda \,x} \,i \,
 \frac{\partial}{\partial \lambda} \,\langle \bar{\psi} (0) \,\gamma^i \,
 {\cal L} [0,\lambda] \,\psi (\lambda) \rangle 
 = - \,i \,\epsilon^{ij}_\perp \,\Delta^j_\perp \,s^+ \ \ \ \ \  
 \nonumber \\
 &\,& \hspace{3mm} \times \ 
 x \,[\,H^q (x,0,0) + E^q (x,0,0) + G^q_2 (x,0,0) \,]
 + \cdots .
\end{eqnarray}
On the other hand, the 3rd term gives
\begin{eqnarray}
 &\,& i \,\epsilon^{i + \beta \rho} \,\int \,\frac{d \lambda}{2 \,\pi} \,
 e^{\,i \,\lambda \,x} \,\,\langle \bar{\psi} (0) \,\gamma_\rho \,\gamma_5 \,
 {\cal L} [0,\lambda] \,D_\beta (\lambda) \,\psi (\lambda) \,\rangle
 \nonumber \\
 &\,& \hspace{4mm} = \ i \,\epsilon^{ij}_\perp \,\Delta^j_\perp \,
 s^+ \,\,\Delta q(x) \ + \ \cdots, \ \ \ \ 
\end{eqnarray}
where $\Delta q(x)$ is the familiar longitudinally polarized PDF of quarks.
Particularly important for our discussion here is the 2nd term.
One can show that \cite{HMS04,HY12}
\begin{eqnarray}
 &\,& \int \,\frac{d \lambda}{2 \,\pi} \,e^{\,i \,\lambda x} \,\,
 \langle \bar{\psi} (0) \,\gamma^+ \,D^i (\lambda) \,\psi (\lambda) \rangle
 \nonumber \\
 &\,& = \ i \,\epsilon^{ij}_\perp \,\Delta^j_\perp \,s^+ \,
 \int \,d x^\prime \,\Phi_D (x,x^\prime).
\end{eqnarray}
Here,  $\Phi_D (x,x^\prime)$ is the $D$-type quark-gluon correlation function
defined by
\begin{eqnarray}
 &\,& \int \,\frac{d \lambda}{2 \,\pi} \int \,\frac{d \mu}{2 \,\pi} \,\,
 e^{\,i \,\lambda \,x} \,e^{\,\mu \,(x^\prime - x)} \nonumber \\
 &\,& \hspace{4mm} \times \ \langle p^\prime,s^\prime \,|\,
 \bar{\psi} (0) \,\gamma^+ \,{\cal L} [0,\mu] \,D^i (\mu) \,{\cal L} [\mu, \lambda] \,
 \psi (\lambda) \,|\,p,s \rangle \ \ \ \nonumber \\
 &\,& \hspace{5mm} = \ \epsilon^{+ i \rho \sigma} \,
 \bar{s}_\rho \,\Delta_{\perp,\sigma} \,
 \Phi_D (x,x^\prime) \ + \ \cdots. \ \ \ \ \  
\end{eqnarray}
Collecting these three terms, and integrating over $x$, we are thus led to the
relation
\begin{equation}
 2 \,J^q \ + \ \int \,x \,G^q_2 (x,0,0) \,d x \ - \ L^q \ - \ \Delta q \ = \ 0.
\end{equation}
Here, use has been made of the familiar Ji relation,
\begin{equation}
 \int \,x \,[\,H^q (x,0,0) \ + \ E^q (x,0,0) \,] \, d x \ = \ 2 \,J^q ,
\end{equation}
together with the obvious relation,
\begin{equation}
 \int \,d x \int \,d x^\prime \,\,\Phi_D (x,x^\prime) \ = \ L^q .
\end{equation}
Since $2 \,J^q = 2 \,L^q + \Delta q$, it then follows that
\begin{equation}
 L^q \ = \ - \,\int \,x \,G^q_2 (x,0,0) \,d x
\end{equation}
It is very important to recognize the fact that the quark OAM appearing in the
above manipulation is the manifestly gauge-invariant mechanical OAM not
the canonical OAM. This is perfectly consistent with our general
statement before that what appears in the equation of motion is the mechanical
OAM not the canonical OAM.

Although the quark OAM appearing in the above discussion is the mechanical
OAM, it is not impossible to divide it into the canonical OAM and the potential
OAM, as was done in the analysis of Hatta and Yoshida \cite{HY12}.
This uses the relation between the D-type and F-type quark-gluon correlation
functions elaborated in the work by Eguchi, Koike, and Tanaka \cite{EKT06}.
Using their technique, it can be shown that
the F-type quark-gluon correlation function $\Phi_F (x,x^\prime)$ defined
by the equation,
\begin{eqnarray}
 &\,& \int \,\frac{d \lambda}{2 \,\pi} \int \,\frac{d \mu}{2 \,\pi} \,\,
 e^{\,i \,\lambda \,x} \,e^{\,i \,\mu \,(x^\prime - x)} \nonumber \\
 &\,& \times \ \ \langle p^\prime,s^\prime \,|\,
 \bar{\psi} (0) \,\gamma^+ \,{\cal L} [0,\mu] \,g \,F^{+i} (\mu) \,
 {\cal L} [\mu, \lambda] \,\psi (\lambda) \,|\,p,s \rangle \nonumber \\
 &\,& = P^+ \,\epsilon^{ij}_\perp \,s^j_\perp \,\Psi_F (x,x^\prime) +  
 \epsilon^{ij}_\perp \,\Delta^j_\perp \,s^+ \,\Phi_F (x,x^\prime) + \cdots ,
 \ \ \ \ \ \ \ \ 
\end{eqnarray}
has the symmetry
\begin{equation}
 \Phi_F (x,x^\prime) \ = \ - \,\Phi_F (x^\prime, x) ,
\end{equation}
and that it is not independent of the $D$-type quark-gluon correlation
function $\Phi_D (x,x^\prime)$, which we know is related to the mechanical OAM.
In fact, Hatta and Yoshida showed that the following relation holds \cite{HY12}
\begin{equation}
 \Phi_D (x,x^\prime) \ = \ {\cal P} \,\frac{1}{x - x^\prime} \,
 \Phi_F (x,x^\prime) \ + \ 
 \delta (x - x^\prime) \,L^q_{can} (x).
\end{equation}
Here, $L^q_{can} (x)$ is the canonical OAM distribution of quarks, whose integral
over $x$ gives the net canonical OAM. We are interested here only in the net OAM,
because the local distribution can easily be altered by the surface terms
usually neglected in the nucleon spin decomposition problem.
(The full function of surface terms in the nucleon spin decomposition
problem still remains to be clarified \cite{Lowdon14,LM15}.)
From the above relation, we thus obtain
\begin{equation}
 L^q_{can} \ = \ L^q_{mech} \ + \ L^q_{pot} , 
\end{equation}
with
\begin{eqnarray}
 L^q_{mech} &=& \int \,dx \int \,d x^\prime \,\,\Phi_D (x,x^\prime), \\
 L^q_{pot} &=& \int \,d x \int \,d x^\prime \,\,{\cal P} \,\,\frac{1}{x^\prime - x} \,\,
 \Phi_F (x,x^\prime) . \ \ \ 
\end{eqnarray}
As pointed out before, the mechanical OAM $L^q_{mech}$ can be related to
an observable, i.e. the GPD $G^q_2$.
On the other hand, in order to know $L^q_{can}$ or $L^q_{pot}$, we need to
know the full information of the $F$-type quark-gluon correlation function, or
at least its principle-part integral. It is not clear whether the extraction
of it can be done in a theoretical-scheme-independently way or not.

In a recent paper \cite{Lorce15}, Lorce carried out an exhaustive analysis
of the matrix elements of various forms of energy-momentum tensors as
well as the related angular momentum tensors, aiming at clarifying the relation
between several forms of nucleon spin decomposition.
He concluded that two-parton TMD distributions cannot provide any
qualitative model-independent information about the parton OAM.
This statement is consistent with our conclusion that the GTMD $F_{14}$
does not appear in any established factorization schemes.
Among several relations derived in his paper, particularly relevant to
our discussion is the relation showing the difference between the
canonical OAM and mechanical (or kinetic) OAMs of quarks and gluons : 
\begin{eqnarray}
 \langle L^q \rangle_{can} \ - \ \langle L^q \rangle_{mech} &=& 
 - \,[\,\langle L^G \rangle_{can} \ - \ \langle L^G \rangle_{mech} \,]
 \ \ \ \ \ \nonumber \\
 &=& \ s^+ \,A^{e,3}_4 (0,0) .
\end{eqnarray}
Here, $A^{e,3}_4 (\xi,t)$ is the time-even part of energy-momentum form
factor defined by the following Lorentz structure of the nucleon matrix
element of the tensor $T^{\mu \nu}_3$ : 
\begin{eqnarray}
 &\,& \langle p^\prime, s^\prime \,|\, T^{\mu \nu}_3 (0) \,|,\,p, s \rangle
 \nonumber \\
 &\,& = \ 
 \bar{u} (p^\prime, s^\prime) \,
 \frac{P^\mu \,i \,\sigma^{\nu \rho} \,\Delta_\rho}{2 \,M_N} \,
 A^{e,3}_4 (\xi,t) \,u (p,s) , \ \ \ 
\end{eqnarray}
where
\begin{equation}
 T^{\mu \nu}_3 (x) \ = \ - \,\bar{\psi} (x) \,\gamma^\mu \,
 g \,A^\nu_{phys} (x) \, \psi (x) .
\end{equation}
Undoubtedly, aside from overall sign difference, $A^{e,3}_4 (0,0)$ just
corresponds to the potential angular momentum,
which is related to the twist-3 quark-gluon correlation function
$\Phi_F (x,x^\prime)$ as
\begin{equation}
 A^{e,3} (0,0) \ = \ - \,
 \int \,d x \int \,d x^\prime \,\,{\cal P} \,\frac{1}{x^\prime - x}
 \,\,\Phi_F (x,x^\prime). 
\end{equation}
This energy-momentum form factor, corresponding to the tensor
operator containing only the physical component of the gluon field, does
not appear in the cross-section formula of the standard GPD factorization
scheme. Again, this would throw some doubt on the direct
observability of the canonical OAM. 
It should be contrasted to the fact that the GPD $G_2$,
the 2nd moment of which gives the mechanical quark OAM, appears in
the GPD factorization formula, so that it can in principle be thought of
as an observable. Naturally, to extract the twist-3 GPD $G_2$ would not
be a practically  easy task. However, we recall the fact that, although
indirectly, the mechanical quark OAM can be accessed already at the
twist-2 level through the relation \cite{Ji97} :
\begin{equation}
 L^q_{mech} =  \frac{1}{2} \,\int  x \,\left[
 H^q (x,0,0) + E^q (x,0,0) \right] \, dx \, - \,
 \frac{1}{2} \,\Delta q . 
\end{equation}
In principle, a similar analysis is possible also for the gluon mechanical
OAM \cite{Waka11A}, although the extraction of the gluon GPDs is
extremely hard even at the twist-2 level.

\section{Summary and Conclusion}
\label{sec:4}

We have investigated the problem of the gauge-invariant complete
decomposition of the nucleon spin. The central question here is
whether the total gluon angular momentum can be decomposed
into its spin and orbital parts without conflicting the
gauge-invariance principle.
Although many recent investigations show that the answer is certainly
affirmative, no clear answer has ever been given as to the
question whether it does not contradict the standard textbook statement
that the total angular momentum of a massless particle like the
photon or the gluon cannot be gauge-invariantly decomposed into its
spin and orbital parts. We have shown that what enables to circumvent
the conflict is the existence of a particular spatial direction in
the physics that we are dealing with. It is the direction of the
nucleon momentum in the parton physics, which plays the same role as
the direction of paraxial laser beam in the physics of photon spin and OAM
measurements. The physical or transverse components of the gluon or
the photon, which are respectively gauge-covariant and gauge-invariant,
can consistently be defined with respect to these axises, even though the
concept of {\it transversality} is in general Lorentz-frame dependent.

The uniqueness or nonuniqueness problem of the gluon spin or the
gauge-invariant complete decomposition of the nucleon spin can also
be easily resolved once we notice the existence of this particular
direction in the parton physics. We have shown that the physical
requrement of the Lorentz-boost invariance along the direction of the
parent nucleon momentum selects one favorable decomposition of the gluon
spin from many possible candidates. It is the definition based on
the light-cone gauge motivated choice of the physical component
of the gluon. This means that, in some sense, the
Lorentz symmetry plays more vital role than the gauge symmetry
in the unambiguous definition of the gluon spin, which can be probed
by the DIS measurements.

We have also discussed another practically more important problem
of the nucleon spin decomposition, i.e. the relation with direct
observables. Of particular interest to us is to unravel relative
merits of the canonical-type decomposition and the
mechanical-type one from the observational viewpoint.
Now that both decomposition satisfy gauge-invariance, the gauge
principle cannot say anything about the superiority and
inferiority of these decompositions.
What is a crucial difference between these two OAMs from the
physical point of view, then ?
In our opinion, a superiority of the mechanical OAM over the
canonical one lies in the fact that the former not the latter
appears in the equation of motion with the Lorentz force.
In fact, we have clarified the fact that the QCD equation of motion
plays an important role in establishing the connection between
the mechanical quark OAM and the observable GPD $G_2$,
which appears in the collinear (GPD) factorization scheme. 

We have also verified non-universal nature of the statement that
the momentum variable in the Wigner distribution refers to the
canonical momentum not the mechanical momentum.
Instead, we gave a universally correct physical interpretation of
the Wigner-distribution-motivated definitions of the average transverse
momentum as well as the average longitudinal OAM.
It is true that the average longitudinal OAM of quarks defined by the
Wigner distribution coincides with the canonical OAM not the
mechanical OAM. The physical reason of it can be easily understood
once one accepts our interpretation that the average OAM defined
by the Wigner distribution represents the OAM of a quark in the
asymptotic region after leaving the spectator in the semi-inclusive DIS
reaction. Note that it is the OAM of a free quark if it were not for the
color confinement.
Unfortunately, this asymptotic OAM of a quark is not a direct observable
because of the color confinement of QCD. The fact is that, any practical
extraction through DIS measurements must be based on the known
TMD or collinear factorization scheme, and that we do not know yet
any TMD or GPD related to the canonical OAM.   

\vspace{3mm}

\begin{acknowledgement}
The author greatly acknowledges enlightening discussion with 
E.~Leader, M.~Anselmino, and C.~Lorc\'{e} during the ECT* workshop
on "Spin and Orbital Angular Momentum of Quarks and Gluons in the
Nucleon". He also appreciate many helpful discussions with K. Tanaka.
Several useful comments by S.~C.~Tiwari are also greatly acknowledged.
\end{acknowledgement}





\begin{thebibliography}{}

\bibitem{EMC88}
J.~Aschman {\it et al.} (EMC Collaboration) Phys. Lett. B {\bf 206},
364 (1988).

\bibitem{EMC89}
J.~Aschman {\it et al.} (EMC Collaboration ) Nucl. Phys. B {\bf 328},
1 (1989).

\bibitem{JM90}
R. L.~Jaffe and A.~Manohar, Nucl. Phys. B {\bf 337}, 509 (1990).

\bibitem{Review_LL14}
E.~Leader and C.~Lorc\'{e}, Phys. Rep. {\bf 541}, 163 (2014).

\bibitem{Review_Waka14}
M.~Wakamatsu, Int. J. Mod. Phys. A {\bf 29}, 1430012 (2014).

\bibitem{Ji97}
X.~Ji, Phys. Rev. Lett. {\bf 78}, 610 (1997).

\bibitem{Ji98}
X.~Ji, Phys. Rev. D {\bf 58}, 056003 (1998).

\bibitem{Book_AB65}
A. I.~Akhiezer and V. B.~Berestetskii, {\it Quantum Electrodynamics}
(John Willey \& Sons Inc., 1965)

\bibitem{Book_JR76}
J. M.~Jauch and F.~Rohrlich, {\it The Theory of Photons and Electrons}
(Springer-Verlar, Berlin, 1976)

\bibitem{Book_BLP82}
V. B.~Berestetskii, E. M.~Lifshitz and L. P.~Pitaevskii,
{\it Quantum Electrodynamics} (Pergamon, Oxford, 1982)

\bibitem{Book_CDG89}
C.~Cohen-Tannoudji, J.~Dupont-Roc and G.~Grynberg,
{\it Photons and Atoms} (John Wiley \& Sons Inc., 1989)

\bibitem{Beth36}
R. A.~Beth, {\it Phys. Rev.} {\bf 50}, 115 (1936).

\bibitem{ABSW92}
L.~Allen, M. W.~Beijersbergen, R. J. C.~Spreeuw and J. P.~Woerdman,
{\it Phys. Rev.} A {\bf 45}, 8185 (1992).

\bibitem{Chen08}
X.-S.~Chen, X.-F.~L\"{u}, W.-M.~Sun, F.~Wang, and T.~Goldman,
Phys. Rev. Lett. {\bf 100}, 232002 (2008).

\bibitem{Chen09}
X.-S.~Chen, W.-M.~Sun, X.-F.~L\"{u}, F.~Wang, and T.~Goldman,
Phys. Rev. Lett. {\bf 103}, 062001 (2009)

\bibitem{Waka10}
M.~Wakamatsu, Phys. Rev. D {\bf 81}, 114010 (2010).

\bibitem{Waka11A}
M.~Wakamatsu, Phys. Rev. D {\bf 83}, 014012 (2011)

\bibitem{Waka12}
M.~Wakamatsu, Phys. Rev. D {\bf 85}, 114039 (2012).

\bibitem{BJ99}
S. V.~Bashinsky and R. L.~Jaffe, Nucl. Phys. B {\bf 536}, 303 (1999).

\bibitem{JXY12}
X.~Ji, X.~Xiong and F.~Yuan, 
{\it Phys. Rev. Lett.} {\bf 109}, 152005 (2012).

\bibitem{JXZ12}
X.~Ji, Y.~Xu and Y.~Zhao, 
{\it JHEP} {\bf 1208}, 082 (2012).

\bibitem{Lorce13A}
C.~Lorc\'{e}, Phys. Rev. D {\bf 87}, 034031(2013).

\bibitem{Lorce13B}
C.~Lorc\'{e}, Phys. Lett. B {\bf 719}, 185 (2013).

\bibitem{Lorce14}
C.~Lorc\'{e}, Nucl. Phys. A {\bf 925}, 1 (2014).

\bibitem{Waka13}
M.~Wakamatsu, Phys. Rev. D {\bf 87}, 094035 (2013).

\bibitem{Stoilov14}
N.~Stoilov, Bulg. J. Phys. {|bf 41}, 251 (2014).

\bibitem{FLM14}
J.~Franc\c{o}is, S.~Lazzarini and T.~Masson, 
Phys. Rev. D {\bf 91}, 045014 (2015).

\bibitem{JZZ13}
X.~Ji, J. H.~Zhang and Y.~Zhao, Phys. Rev. Lett. {\bf 111}, 112002 (2013). 

\bibitem{Book_Collins11}
J.C.~Collins, {\it Foundations of Perturbative QCD} \\
(Cambridge University Press, Cambridge, 2011).

\bibitem{HJZ14}
Y.~Hatta, X.~Ji and Y.~Zhao, Phys. Rev. D {\bf 89}, 085030 (2014).

\bibitem{Manohar91}
A. V.~Manohar, Phys. Rev. Lett. {\bf 66}, 289 (1991).

\bibitem{KT99}
J.~Kodaira and K.~Tanaka, Prog. Theor. Phys. {\bf 101}, 191 (1999).

\bibitem{Hatta11}
Y.~Hatta, Phys. Rev. D {\bf 84}, 041701 (2011).

\bibitem{VanEN94A}
S. J.~Van Enk and G.~Nienhuis, {\it J. Mod. Opt.} {\bf 41}, 963 (1994).

\bibitem{VanEN94B}
S. J.~Van Enk and G.~Nienhuis, {\it Europhys. Lett.} {\bf 25}, 497 (1994).

\bibitem{BA94}
S. M.~Barnett and L.~Allen, 1994 Opt. Commun. {\bf 110}, 670 (1994).

\bibitem{BAOA10}
K. Y.~Bliokh, M. A.~Alonso, E. A.~Ostrovskaya and A.~Aiello,
Phys. Rev. A {\bf 82}, 063825 (2010).

\bibitem{Leader13}
E.~Leader, 2013 Phys. Lett. B {\bf 720}, 120 (2013).

\bibitem{HTY12}
Y.~Hatta, K.~Tanaka and S.~Yoshida, JHEP {\bf 1210}, 080 (2012).

\bibitem{HKM14}
A.~Harindranath, R.~Kundu and A.~Mukherjee, Phys. Lett. B {\bf 728}, 63 (2014).

\bibitem{Jaffe96}
R. L.~Jaffe, Phys. Lett. B {\bf 365}, 359 (1996).

\bibitem{Berger99}
M. A.~Berger, Plasma Phys. Control. Fusion {\bf 41}, B167 (1999).

\bibitem{Burkardt13}
M.~Burkardt, Phys. Rev. D {\bf 88}, 14014 (2013).

\bibitem{Hatta12}
Y.~Hatta, Phys. Lett. B {\bf 708}, 186 (2012).

\bibitem{Collins02}
J. C.~Collins, Phys. Lett. B {\bf 536}, 43 (2002).

\bibitem{BJY03}
A. V.~Belitsky, X.~Ji and F.~Yuan, Nucl. Phys. B {\bf 656}, 165 (2003).

\bibitem{BMP03}
D.~Boer, P. J.~Mulders and F.~Pijlman, Nucl. Phys. B {\bf 667}, 201 (2003).

\bibitem{ET85}
A. V.~Efremov and O. V.~Teryaev, Phys. Lett. B {\bf 150}, 383 (1985).

\bibitem{QS91}
J.~Qiu and G.~Sterman, Phys. Rev. D {\bf 59}, 014004 (1991).

\bibitem{QS92}
J.~Qiu and G.~Sterman, Nucl. Phys. B {\bf 378}, 52 (1992).

\bibitem{MMS09}
S.~Meissner, A.~Metz and M.~Schlegel, JHEP {\bf 0908}, 056 (2009).

\bibitem{Sivers90}
D.~Sivers, Phys. Rev. D {\bf 41}, 83 (1990).

\bibitem{Sivers91}
D.~Sivers, Phys. Rev. D {\bf 43}, 261 (1991).

\bibitem{HMS04}
Ph.~Haegler, A.~Mukherjee and A.~Schaefer, Phys. Lett. B {\bf 582},
55 (2004).

\bibitem{HY12}
Y.~Hatta and S.~Yoshida, JHEP {\bf 1210}, 080 (2012).

\bibitem{LP11}
C.~Lorc\'{e} and B.~Pasquini, Phys. Rev. D {\bf 84}, 014015 (2011).

\bibitem{PPSS00}
M.~Penttinen, M. V.~Polyakov, A. G.~Shuvaev and M.~Strikman, 2000
Phys. Lett. B {\bf 491}, 96 (2000).

\bibitem{KP04}
D. V.~Kiptily and M.V.~Polyakov, Eur. Phys. J. C {\bf 37}, 105 (2004).

\bibitem{CGHLR14A}
A.~Courtoy, G. R.~Goldstein, J. O. G.~Hernandez, S.~Liuti and A.~Rajan \\
Phys. Lett. B {\bf 731}, 141 (2014)

\bibitem{CGHLR14B}
A.~Courtoy, G. R.~Goldstein, J. O. G.~Hernandez, S.~Liuti and A.~Rajan \\
arXiv : 1412.0647 [hep-ph]

\bibitem{Amado79}
R. D.~Amado, Phys. Rev. C {\bf 19}, 1473 (1979).

\bibitem{Friar79}
J. L.~Friar, Phys. Rev. C {\bf 20}, 325 (1979).

\bibitem{BFRS07}
S. K.~Bogner, R. J.~Furnstahl, S.~Ramanan and A.~Schwenk,
Nucl Phys. A {\bf 784}, 79 (2007).

\bibitem{Ratcliffe86}
P. G.~Ratcliffe, Nucl Phys. B {\bf 264}, 493 (1986).

\bibitem{BB91}
I. I.~Balitsky and V. M.~Braun, Nucl Phys. B {\bf 361}, 93 (1991).

\bibitem{EKT06}
H.~Eguchi, Y.~Koike and K.~Tanaka, Nucl. Phys. B {\bf 752}, 1 (2006).

\bibitem{Lowdon14}
P.~Lowdon, Nucl. Phys. B {\bf 889}, 801 (2014).

\bibitem{LM15}
T.~Liu and B.-Q.~Ma, Phys. Rev. D {\bf 91}, 017501 (2015).

\bibitem{Lorce15}
C.~Lorce\'{e}, arXiv : 1502.06656 [hep-ph] 


\end{thebibliography}

\end{document}